\documentclass[twocolumn]{aastex63}

\usepackage{natbib}
\usepackage{enumerate}
\usepackage{graphicx}
\usepackage{epstopdf}
\usepackage{amsmath}
\usepackage{amssymb}
\usepackage{comment}
\usepackage{bm}
\usepackage{color}
\usepackage{multirow}
\usepackage{booktabs, makecell, multirow}
\usepackage{rotating}

\newcommand{\aref}{$^{(a)}$}
\newcommand{\bref}{$^{(b)}$}
\newcommand{\cref}{$^{(c)}$}
\newcommand{\dref}{$^{(d)}$}
\newcommand{\eref}{$^{(e)}$}
\newcommand{\fref}{$^{(f)}$}
\newcommand{\gref}{$^{(g)}$}
\newcommand{\mbh}{$M_{\rm BH}$}

\newcommand{\Hb}{H$\beta$ }
\newcommand{\logfmeansigma}{$\log_{10}(f_{{\rm mean},{\sigma}})$}
\newcommand{\logfrmssigma}{$\log_{10}(f_{{\rm rms},{\sigma}})$}
\newcommand{\logfrmsfwhm}{$\log_{10}(f_{{\rm rms},{\rm FWHM}})$}
\newcommand{\logfmeanfwhm}{$\log_{10}(f_{{\rm mean},{\rm FWHM}})$}
\newcommand{\lineprofile}{$\log_{10}(\rm{FWHM}/\sigma)_{\rm{rms}}$}
\newcommand{\mrka}{Mrk 1392 }
\newcommand{\pga}{PG 2209+184 }
\newcommand{\rbsa}{RBS 1917 }

\newcommand{\rbsb}{RBS 1303 }
\newcommand{\mcga}{MCG +04-22-042 }

\newcommand{\rxja}{RXJ 2044.0+2833}
\newcommand{\mrke}{Mrk 841}
\newcommand{\three}{3C 382}

\defcitealias{pancoast14b}{P14}
\defcitealias{Grier++17}{G17}
\defcitealias{2018ApJ...866...75W}{W18}
\defcitealias{2020ApJ...902...74W}{W20}
\defcitealias{bentz2021detailed}{B21}
\defcitealias{pancoast11}{P11}
\defcitealias{Villafa_a_2022}{V22}
\defcitealias{2022ApJ...934..168B}{B22}

\shorttitle{What Does the Geometry of the \Hb BLR Depend On?}
\shortauthors{Villafa\~na et al.}
 
\begin{document}
\title{What Does the Virial Coefficient of the \Hb Broad-Line Region Depend On?}
\author[0000-0002-1961-6361]{Lizvette Villafa\~na}
\affiliation{Department of Physics and Astronomy, University of California, Los Angeles, CA 90095-1547, USA}

\author[0000-0002-4645-6578]{Peter R. Williams}
\affiliation{Department of Physics and Astronomy, University of California, Los Angeles, CA 90095-1547, USA}

\author[0000-0002-8460-0390]{Tommaso Treu}
\affiliation{Department of Physics and Astronomy, University of California, Los Angeles, CA 90095-1547, USA}

\author{Brendon J. Brewer}
\affiliation{Department of Statistics, The University of Auckland, Private Bag 92019, Auckland 1142, New Zealand}

\author[0000-0002-3026-0562]{Aaron J. Barth}
\affiliation{Department of Physics and Astronomy, 4129 Frederick Reines Hall, University of California, Irvine, CA 92697, USA}

\author[0000-0002-1912-0024]{Vivian U}
\affiliation{Department of Physics and Astronomy, 4129 Frederick Reines Hall, University of California, Irvine, CA 92697, USA}
\affiliation{Department of Physics and Astronomy, University of California, Riverside, CA 92521, USA}

\author[0000-0003-2064-0518]{Vardha N. Bennert}
\affiliation{Physics Department, California Polytechnic State University, San Luis Obispo CA 93407, USA}


\author[0000-0001-8416-7059]{Hengxiao Guo}
\affiliation{Department of Physics and Astronomy, 4129 Frederick Reines Hall, University of California, Irvine, CA 92697, USA}
\affiliation{Key Laboratory for Research in Galaxies and Cosmology, Shanghai Astronomical Observatory, Chinese Academy of Sciences, 80 Nandan Road, Shanghai 200030, People’s Republic of China}
\affiliation{Department of Physics and Astronomy, 4129 Frederick Reines Hall, University of California, Irvine, CA, 92697-4575, USA}

\author[0000-0002-2816-5398]{Misty C. Bentz}
\affiliation{Department of Physics and Astronomy, Georgia State University, Atlanta, GA 30303, USA}

\author[0000-0003-4693-6157]{Gabriela Canalizo}
\affiliation{Department of Physics and Astronomy, University of California, Riverside, CA 92521, USA}

\author[0000-0003-3460-0103]{Alexei V. Filippenko}
\affiliation{Department of Astronomy, University of California, 501 Campbell Hall, Berkeley, CA 94720-3411, USA}

\author[0000-0002-3739-0423]{Elinor Gates}
\affiliation{Lick Observatory, P.O. Box 85, Mt. Hamilton, CA 95140, USA}


\author[0000-0003-0634-8449]{Michael D. Joner}
\affiliation{Department of Physics and Astronomy, N283 ESC, Brigham Young University, Provo, UT 84602, USA}

\author{Matthew A. Malkan}
\affiliation{Department of Physics and Astronomy, University of California, Los Angeles, CA 90095-1547, USA}

\author[0000-0002-8055-5465]{Jong-Hak Woo}
\affil{Astronomy Program, Department of Physics and Astronomy, Seoul National University, 1 Gwanak-ro, Gwanak-gu, Seoul 08826, Korea}
\affil{SNU Astronomy Research Center, Seoul National University, 1 Gwanak-ro, Gwanak-gu, Seoul 08826, Republic of Korea}

\author{Bela Abolfathi}
\affiliation{Department of Physics and Astronomy, 4129 Frederick Reines Hall, University of California, Irvine, CA 92697, USA}


  

\author{Thomas Bohn}
\affiliation{Department of Physics and Astronomy, University of California, Riverside, CA 92521, USA}


\author[0000-0002-4924-444X]{K. Azalee Bostroem}
\affiliation{DiRAC Institute, Department of Astronomy, University of Washington, 3910 15th Avenue, NE, Seattle, WA 98195, USA}

\author{Andrew Brandel}
\affiliation{Department of Physics and Astronomy, 4129 Frederick Reines Hall, University of California, Irvine, CA 92697, USA}

\author[0000-0001-5955-2502]{Thomas G. Brink}
\affiliation{Department of Astronomy, University of California, 501 Campbell Hall, Berkeley, CA 94720-3411, USA}

\author{Sanyum Channa}
\affiliation{Department of Physics, University of California, Berkeley, CA 94720, USA}
\affiliation{Department of Physics, Stanford University, Stanford, CA 94305, USA}


\author[0000-0002-2248-6107]{Maren Cosens}
\affiliation{Physics Department, University of California, San Diego, 9500 Gilman Drive, La Jolla, CA 92093 USA}
\affiliation{Center for Astrophysics and Space Sciences, University of California, San Diego, 9500 Gilman Drive, La Jolla, CA 92093 USA}

\author{Edward Donohue}
\affiliation{Physics Department, California Polytechnic State University, San Luis Obispo CA 93407, USA}
\affiliation{Booz Allen, 1615 Murray Canyon Road, Suite 8000, San Diego, CA 92108, USA}



\author[0000-0002-7232-101X]{Goni Halevi}
\affiliation{Department of Astronomy, University of California, 501 Campbell Hall, Berkeley, CA 94720-3411, USA}
\affiliation{Department of Astrophysical Sciences, Princeton University, 4 Ivy Lane, Princeton, NJ 08544, USA}


\author[0000-0003-0034-5909]{Carol E. Hood}
\affiliation{Department of Physics, California State University, San Bernardino, 5500 University Parkway, San Bernardino, CA 92407, USA}


\author{J. Chuck Horst}
\affiliation{Department of Astronomy, San Diego State University, San Diego, CA 92182-1221, USA}

\author{Maxime de Kouchkovsky}
\affiliation{Department of Astronomy, University of California, 501 Campbell Hall, Berkeley, CA 94720-3411, USA}

\author{Benjamin Kuhn}
\affiliation{Space Telescope Science Institute, 3700 San Martin Drive, Baltimore, MD 21218, USA}
\affiliation{Department of Astronomy, San Diego State University, San Diego, CA 92182-1221, USA}


\author[0000-0001-7839-1986]{Douglas C. Leonard}
\affiliation{Department of Astronomy, San Diego State University, San Diego, CA 92182-1221, USA}




\author[0000-0003-1263-808X]{Ra\'ul Michel}
\affiliation{Instituto de Astronom\'ia, Universidad Nacional Aut\'onoma de M\'exico, AP 877, Ensenada, Baja California, C.P. 22830 M\'exico}

\author{Melanie Kae B. Olaes}
\affiliation{Department of Astronomy, San Diego State University, San Diego, CA 92182-1221, USA}
  
\author[0000-0001-9877-1732]{Daeseong Park}
\affiliation{Department of Astronomy and Atmospheric Sciences, Kyungpook National University, Daegu, 41566, Republic of Korea}
\affiliation{Korea Astronomy and Space Science Institute, Daejeon, 34055, Republic of Korea}




\author[0000-0003-4852-8958]{Jordan N. Runco}
\affiliation{Department of Physics and Astronomy, University of California, Los Angeles, CA 90095-1547, USA}



\author[0000-0003-3432-2094]{Remington O. Sexton}
\affiliation{Department of Physics and Astronomy, University of California, Riverside, CA 92521, USA}
\affiliation{U.S. Naval Observatory, 3450 Massachusetts Ave NW, Washington, DC 20392-5420, USA}
\affiliation{Department of Physics and Astronomy, George Mason University, 4400 University Dr, Fairfax, VA 22030-4444, USA}


\author{Isaac Shivvers}
\affiliation{Department of Astronomy, University of California, 501 Campbell Hall, Berkeley, CA 94720-3411, USA}

\author[0000-0002-4202-4188]{Chance L. Spencer}
\affiliation{Physics Department, California Polytechnic State University, San Luis Obispo CA 93407, USA}
\affiliation{Department of Physics, California State University Fresno, Fresno, CA 93740-8031, USA}

\author[0000-0002-3169-3167]{Benjamin E. Stahl}
\affiliation{Department of Astronomy, University of California, 501 Campbell Hall, Berkeley, CA 94720-3411, USA}
\affiliation{Department of Physics, University of California, Berkeley, CA 94720, USA}

\author{Samantha Stegman}
\affiliation{Department of Astronomy, University of California, 501 Campbell Hall, Berkeley, CA 94720-3411, USA}
\affiliation{Department of Chemistry, University of Wisconsin, Madison, WI 53706, USA}



\author[0000-0002-1881-5908]{Jonelle L. Walsh}
\affiliation{George P. and Cynthia W. Mitchell Institute for Fundamental Physics and Astronomy, Department of Physics \& Astronomy, Texas A\&M University, 4242 TAMU, College Station, TX 77843, USA}


\author{WeiKang Zheng}
\affiliation{Department of Astronomy, University of California, 501 Campbell Hall, Berkeley, CA 94720-3411, USA}

\correspondingauthor{Lizvette Villafa\~na}
\email{lvillafana@astro.ucla.edu}

\begin{abstract} We combine our dynamical modeling black hole mass measurements from the Lick AGN Monitoring Project 2016 sample with measured cross-correlation time lags and line widths to recover individual scale factors, $f$, used in traditional reverberation mapping analyses. We extend our sample by including prior results from Code for AGN Reverberation and Modeling of Emission Lines (\textsc{caramel}) studies that have utilized our methods. Aiming to improve the precision of black hole mass estimates, as well as uncover any regularities in the behavior of the broad-line region (BLR), we search for correlations between $f$ and other AGN/BLR parameters.  We find (i) evidence for a correlation between the virial coefficient \logfmeansigma\ and black hole mass, (ii) marginal evidence for a similar correlation between \logfrmssigma\ and black hole mass, (iii) marginal evidence for an anti-correlation of BLR disk thickness with \logfmeanfwhm and \logfrmsfwhm,
and (iv) marginal evidence for an anti-correlation of inclination angle with \logfmeanfwhm, \logfrmssigma, and \logfmeansigma. Lastly, we find  marginal evidence for a correlation between line-profile shape, when using the root-mean-square spectrum, $\log_{10}(\rm{FWHM}/\sigma)_{\rm{rms}}$, and the virial coefficient, \logfrmssigma, and investigate how BLR properties might be related to line-profile shape using \textsc{caramel} models.
\end{abstract}

\section{Introduction}\label{sec:intro}
It is widely accepted that most galaxies host a supermassive black hole in their center. When the black hole accretes material, it gives rise to a bright central source, known as an active galactic nucleus (AGN). Tight correlations between black hole mass and host-galaxy properties \citep[e.g.,][]{Magorrian++98, ferrarese00, gebhardt00, kormendy13}  
suggest that AGNs play an important role in galaxy evolution.  To understand such a link, both a better understanding of the central regions of AGNs \textit{and} improved black hole mass estimates are needed \citep{ding2020mass}. Black hole mass estimators applicable to cosmologically significant lookback times are particularly desirable as they allow for the determination of the cosmic evolution of the galaxy black hole mass correlations \citep[e.g.,][]{2004ApJ...615L..97T,2006ApJ...645..900W,Salviander_2007,2008ApJ...681..925W,2009ASPC..419..392S,2010ApJ...708.1507B, 2011A&A...535A..87S,Targett_2012, ding2020mass}.

Beyond our local universe, the black hole's gravitational sphere of influence cannot be spatially resolved with current technology, and thus dynamical black hole mass measurements (e.g., modeling stellar/gas kinematics) cannot be constrained \citep[]{kormendy95,ferrarese05}, with rare exceptions \citep[e.g., 3C $273$, IRAS $09149$--$6206$;][]{2018Natur.563..657G, 2020A&A...643A.154G}. Instead, reverberation mapping is the primary tool used to estimate black hole masses in the distant universe, with a limited application to broad-line (Type~1) AGNs.

The technique resolves the gravitational sphere of influence of the central black hole in time by utilizing variations in the continuum that are later reverberated by the broad emission lines (\citealt{blandford82}; \citealt{peterson93}; for a review, see \citealt{2021iSci...24j2557C}). Assuming the delay in variations is due solely to light-travel time, the radius of the broad-line region (BLR) is measured by combining the observed time lag, $\tau$, with the speed of light. A second key assumption --- BLR kinematics are dominated by the black hole's gravity --- provides the velocity of the emitting gas, $v$, as determined by the width of the broad line. Combining the size of the BLR with its velocity, a virial constraint of the black hole's mass (\mbh) is given by

\begin{equation}
    M_{\rm BH} = f \frac{c\tau v^2}{G} = fM_{\rm{vir}},
    \label{eq: rev_map}
\end{equation}
where $f$, or the ``virial coefficient," is a dimensionless scale factor of order unity that captures the relation between measured line-shape parameters and BLR geometry/dynamics, and $c\tau v^2/G$ is referred to as the virial product ($M_{\rm{vir}}$).

In principle, construction of a velocity-delay map, which maps continuum variations to the broad-line flux variations as both a function of line-of-sight velocity and time delay, allows one to constrain the BLR geometry \citep{blandford82}. In practice, however, interpretation is nontrivial, and much about the structure and kinematics of the BLR still remains unknown. For this reason, it is currently impossible to determine the scale factor for an individual AGN using traditional reverberation mapping techniques. Instead, a constant average scale factor, found by aligning reverberation mapped AGNs to the local $M_{\rm{BH}}$--$\sigma_*$ relation, is often used for traditional reverberation mapping black hole mass estimates \citep{onken04, collin06, woo10, woo13, graham11, park12b, grier13b, 2015ApJ...801...38W,Batiste++17}. 

Over the last several years, our team has set out to provide a more reliable way to calibrate the virial coefficient and uncover any regularity in BLR behavior. The discovery of any trends would thus provide \textit{both} insight into the inner regions of AGNs and improve the way black hole masses are calibrated across cosmic time.

Using the methods introduced by \citet{pancoast11}, such as the Code for AGN Reverberation and Modeling of Emission Lines (\textsc{caramel}), we explore a phenomenological description of the BLR and constrain a black hole mass that is consistent with the reverberation mapping dataset, without the need of assuming a scale factor. In this paper, we combine our \textsc{caramel} \mbh\
estimates for the nine sources modeled from the Lick AGN Monitoring Project 2016 \citep[LAMP 2016;][hereafter \citetalias{Villafa_a_2022}]{Villafa_a_2022} with those from prior \textsc{caramel} studies, and determine AGN-specific virial coefficients in order to search for a more reliable way to calibrate $f$.

This paper is organized as follows. We summarize the geometry and kinematics of the \textsc{caramel} model in Section \ref{sec: model_summary} and outline our methodology in calculating AGN-specific virial coefficients in Section~\ref{sec:main_results}. A systematic investigation of correlations between $f$ and observables is carried out in Section~\ref{sec:RESULTS}. Specifically, we consider correlations with  AGN/BLR model parameters in Section~\ref{sec: scale_factor_correlations} and line-profile shape in Section~\ref{sec: scalefactor_lineprofileshape}. We then investigate the effects of BLR geometry and kinematics on BLR line-profile shape in Section \ref{sec: toymodels} and summarize our main conclusions in Section \ref{sec: summary}.

\section{Summary of Relevant \textsc{caramel} Model Parameters} \label{sec: model_summary}
Our work builds on the \textsc{caramel} modeling results of \citetalias{Villafa_a_2022}, \citet[][hereafter \citetalias{2022ApJ...934..168B}]{2022ApJ...934..168B}, \citet[][hereafter \citetalias{bentz2021detailed}]{bentz2021detailed},
\citet[][hereafter \citetalias{2020ApJ...902...74W}]{2020ApJ...902...74W},
\citet[][hereafter \citetalias{2018ApJ...866...75W}]{2018ApJ...866...75W}, \citet[][hereafter \citetalias{Grier++17}]{Grier++17}, and  \citet[][hereafter \citetalias{pancoast14b}]{pancoast14b}. In this section we provide a brief summary of the \textsc{caramel} model detailed by \citetalias{pancoast14b}.

Briefly, \textsc{caramel} is a phenomenological model that uses velocity-resolved reverberation mapping datasets to model the BLR emissivity distribution. The BLR is modeled by point particles, surrounding the black hole located at the origin, which instantaneously reemit light received from the ionizing source, toward an observer.
\subsection{BLR Geometry}
The radial distribution of the BLR point particles is drawn from a Gamma distribution with shape parameter $\alpha$ and scale parameter $\theta$,

\begin{equation}
    p(x|\alpha,\theta)\propto x^{\alpha-1}\exp\Big(-\frac{x}{\theta}\Big)\,.
\end{equation}
\noindent
The distribution is then shifted from the origin by the Schwarzchild radius plus a free parameter $r_{\mathrm{min}}$, which sets the minimum BLR radius. This is then followed by a change of variables from $(\alpha, \theta, r_{\mathrm{min}})$ to $(\mu, \beta, F)$, such that:

\begin{equation}
    \mu=r_{\mathrm{min}}+\alpha \theta \, ,
\end{equation}
\begin{equation} 
\beta=\frac{1}{\sqrt{\alpha}} \, , {~{\rm and}}
\end{equation}
\begin{equation}
F=\frac{r_{\mathrm{min}}}{r_{\mathrm{min}}+\alpha\theta} \, .
\end{equation}
With this change of variables, the two \textsc{caramel} model parameters closely associated with BLR size are $\mu$ and $F$; the parameter $\mu$ describes the mean radius, while the parameter $F$ describes the minimum radius in units of $\mu$.

After the change of variables, the BLR disk thickness is then determined by the model parameter $\theta_o$. The opening angle, $\theta_o$, corresponds to half the angular thickness of the BLR in the angular spherical polar coordinate, such that $\theta_o=90^{\circ}$ corresponds to a spherical BLR. The BLR inclination angle, $\theta_{i}$, is then determined by the angle between a face-on disk and the observer's line of sight. In this way, a face-on BLR geometry would correspond to $\theta_i\rightarrow 0$ and an edge-on BLR geometry $\theta_i\rightarrow 90^{\circ}$.

Once the BLR geometry is determined from the parameters described above and a few additional parameters \citepalias[for a full description of the geometric model, please see][]{pancoast14b}, the kinematics are set by a number of parameters that allow for elliptical, inflowing/outflowing orbits, and macroturbulent contributions.

\subsection{BLR Kinematics}
Particle velocities are modeled using both radial and tangential velocity distributions, with a fraction of particles, $f_{\mathrm{ellip}}$, on near-circular orbits around the central black hole. The remaining $1-f_{\mathrm{ellip}}$ particles can have either inflowing/outflowing orbits, and the direction of motion is determined by the parameter $f_{\mathrm{flow}}$. Inflow motion is defined by values of $f_{\mathrm{flow}}<0.5$  and outflow motion is defined by values of $f_{\mathrm{flow}}>0.5$.

Whether these orbits are bound or unbound is then determined by the parameter $\theta_e$, which describes the angle between escape velocity and circular velocity. In this way, $\theta_e \rightarrow 0^{\circ}$ represents nearly unbound orbits, $\theta_e \rightarrow 90^{\circ}$ represents nearly-circular orbits, and values of $\theta_e\approx45^{\circ}$ represent highly elliptical (bound) orbits. Using the kinematic parameters described above, inflow/outflow motion can be summarized by the $\rm{In.- Out.}$ parameter created by \citetalias{2018ApJ...866...75W},

\begin{equation}
    \rm{In.- Out.} =  \rm{sgn}(\textit{f}_{\mathrm{flow}}-0.5)\times(1-\textit{f}_{\mathrm{ellip}})\times \cos(\theta_e) \, ,
\end{equation}
where sgn is the sign function. Values of $-1$ indicate pure radial inflow and values of $1$ indicate pure radial outflow.

Lastly, in addition to inflow/outflow motion, the model also allows for macroturbulent contributions by including the following $v_{\mathrm{turb}}$ velocity to the line-of-sight velocity:

\begin{equation}
    v_{\mathrm{turb}}=\mathcal{N}(0,\sigma_{
    \mathrm{turb}})|v_{\mathrm{circ}}| \, ,
\end{equation}
where $|v_{\mathrm{circ}}|$ represents circular velocity as determined by the central black hole's mass, and $\mathcal{N}(0,\sigma_{\mathrm{turb}})$ is a Gaussian distribution with standard deviation $\sigma_{\mathrm{turb}}$. The free parameter $\sigma_{\mathrm{turb}}$ is allowed to range from 0.001 to 0.1 and thus represents the contribution of macroturbulent velocities. For each particle, we find the elliptical, inflowing, or outflowing velocity first, and then add the magnitude of the macroturbulent velocity, $v_{\mathrm{turb}}$, determined.

\subsection{Model Results}
In addition to the geometric and dynamical model parameters described above, we also include a black hole mass parameter, \mbh, with a log uniform prior between $2.78\times10^4-1.67\times10^9 M_{\odot}$. Including black hole mass as a model parameter allows us to constrain \mbh, without the use of the scale factor, $f$ (Eqn. \ref{eq: rev_map}). To interpret the results, we use the posterior distribution functions \textsc{caramel} produces for the model parameters, and report the median value and 68\% confidence interval for 1$\sigma$ uncertainties.

In this paper, we use the \textsc{caramel} results found by \citetalias{Villafa_a_2022} for the LAMP 2016 sample, and results from our extended sample's respective papers (\citetalias{pancoast14b, Grier++17, 2018ApJ...866...75W, bentz2021detailed}, and \citetalias{2022ApJ...934..168B}). We note that as outlined in the Appendix of \citet{2022ApJ...935..128W}, the \textsc{caramel} code has undergone some minor modifications since its original publication. These changes were implemented for the work of \citetalias{Villafa_a_2022}, but not for the modeling results of the rest of the subsamples included in our extended sample -- \citetalias{2018ApJ...866...75W}, \citetalias{Grier++17}, \citetalias{pancoast14b}, \citetalias{2020ApJ...902...74W}, \citetalias{bentz2021detailed}, and \citetalias{2022ApJ...934..168B}. However, using a subsample of AGNs modeled with the original code, we have found that the updated code used by \citetalias{Villafa_a_2022} does not significantly change the results produced by the original code (e.g., \citetalias{2018ApJ...866...75W}, \citetalias{Grier++17}, \citetalias{pancoast14b}, \citetalias{2020ApJ...902...74W}, \citetalias{bentz2021detailed}, and \citetalias{2022ApJ...934..168B}) (Colleyn, in prep.). For further details regarding modifications made to the code, please refer to Appendix \ref{caramel modifications}.

\section{The Virial Coefficient} \label{sec:main_results}
A key \textsc{caramel} result is black hole mass, which allows us to determine an AGN-specific virial coefficient for each AGN modeled (see Eqn. \ref{eq: rev_map}). In this section, we summarize the different ways line widths are measured for reverberation mapping black hole mass estimates and our methodology for determining individual AGN virial coefficients. 

\subsection{Line-Width Measurements}
The line width of the broad emission line, which is used to determine the speed of the BLR gas (Eq. \ref{eq: rev_map}), can either be measured from the root-mean-square (rms) spectrum, or from the mean spectrum. Measurements taken from the rms spectrum are computed with the intent that only the variable part of the line will contribute to the line-width calculation \citep{2013BASI...41...61S}. However, whenever the rms profile cannot be measured, owing to insufficient epochs or low signal-to-noise ratio, the line width is often calculated using the mean spectra instead \citep[e.g.,][]{denney10}.

In either case, the line width measured from the spectra selected (i.e., rms or mean) is then characterized by either the full width at half-maximum intensity (FWHM) or the line dispersion, $\sigma_{\rm{line}}$ (i.e., the second moment of the line). The FWHM simply corresponds to the difference between wavelengths from both sides of the peak, $P(\lambda)_{\rm{max}}$, at half of the height. We determine $\sigma_{\rm{line}}$ using the definition of \citet{peterson04}:

\begin{equation}
    \sigma^2_{\rm{line}}(\lambda)=\langle \lambda^2 \rangle - \lambda_0^2 ,
\end{equation}
where 

\begin{equation}
\langle \lambda^2 \rangle = \frac{\int \lambda^2P(\lambda)d\lambda}{\int P(\lambda)d\lambda}
\end{equation}
and 

\begin{equation}
    \lambda_0=\frac{\int \lambda P(\lambda)d\lambda}{\int P(\lambda)d\lambda} .
\end{equation}

Both the width type (i.e., FWHM or $\sigma_{\rm{line}}$) and spectra (i.e., rms or mean) used to measure the line width then determine which calibrated scale factor is needed to calculate the virial \mbh\ (Eq. \ref{eq: rev_map}). For example, \citet{2015ApJ...801...38W} derived a constant $f$ factor based on the $M$--$\sigma_{*}$ relation calibration, for both FWHM-based and $\sigma_{\rm{line}}$-based \mbh\ estimates.

\subsection{AGN-Specific Virial Coefficient Calculations}
For completeness, we determine all four versions of the scale factor ($\log_{10} f_{\rm{FWHM,rms}}$, $ \log_{10} f_{\sigma \rm{,rms}}$,  $\log_{10}f_{\rm{FWHM,mean}}$, and $\log_{10}f_{\sigma \rm{,mean}}$), although measurements using the line dispersion from the rms spectra have been suggested to produce less biased \mbh\ estimates \citep[e.g.,][]{peterson04, 2006A&A...456...75C}.

\subsubsection{LAMP 2016 Sample}
\label{lamp2016_scalefactors}
To calculate the scale factor for each LAMP 2016 source modeled by \citetalias{Villafa_a_2022}, we follow the same approach taken by all other previous \textsc{caramel} works -- we combine the cross-correlation time-lag ($\tau_{\rm{cen}}$) and line-width ($v$) values measured by the campaign's respective reverberation mapping analysis \citep{u2021lick} with the \mbh\ measurements determined from our forward-modelling approach (\mbh\ measurements for the LAMP 2016 sample can be found in \citetalias{Villafa_a_2022}).

To propagate uncertainties, we first assume Gaussian errors on the $\tau_{\rm{cen}}$ and $v$ measurements using the standard deviations listed by \citet{u2021lick} (see Figure \ref{fig:virial_product} below).  For measurements with asymmetrical error bars, the average of the lower and upper errors is used for the standard deviation of the Gaussian distribution.

\begin{figure}
    \centering
    \includegraphics[height=3.0cm, keepaspectratio]{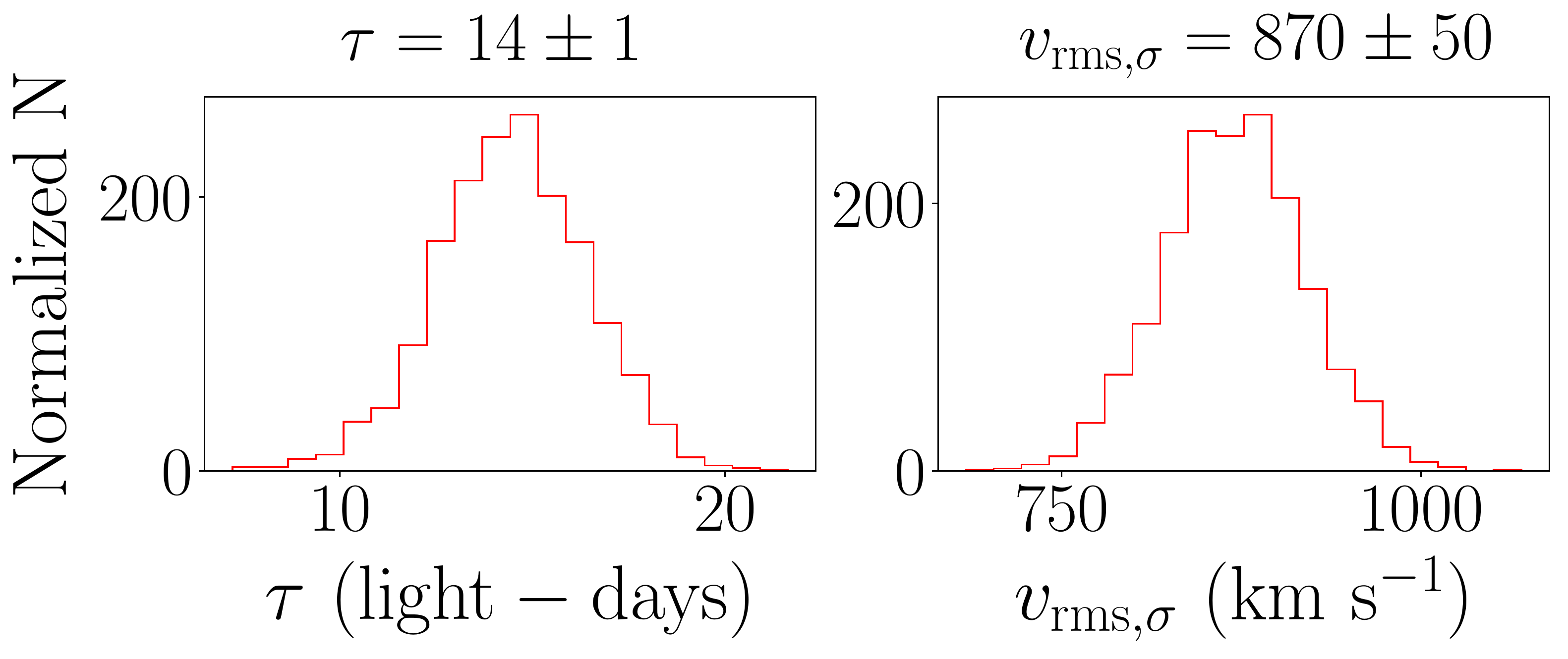}
    \caption{To propagate uncertainties, we assume Gaussian errors on the cross-correlation time lag (left) and line-width (right) measurements given by \citet{u2021lick}. This allows us to create distribution functions that we can utilize with our \textsc{caramel} \mbh\ posterior distribution function to determine the distribution of the scale factor of an individual source, from which we use the 68\% confidence interval for 1$\sigma$ uncertainties.}
    \label{fig:virial_product}
\end{figure}

\begin{figure}
    \centering
    \includegraphics[height=4.0cm, keepaspectratio]{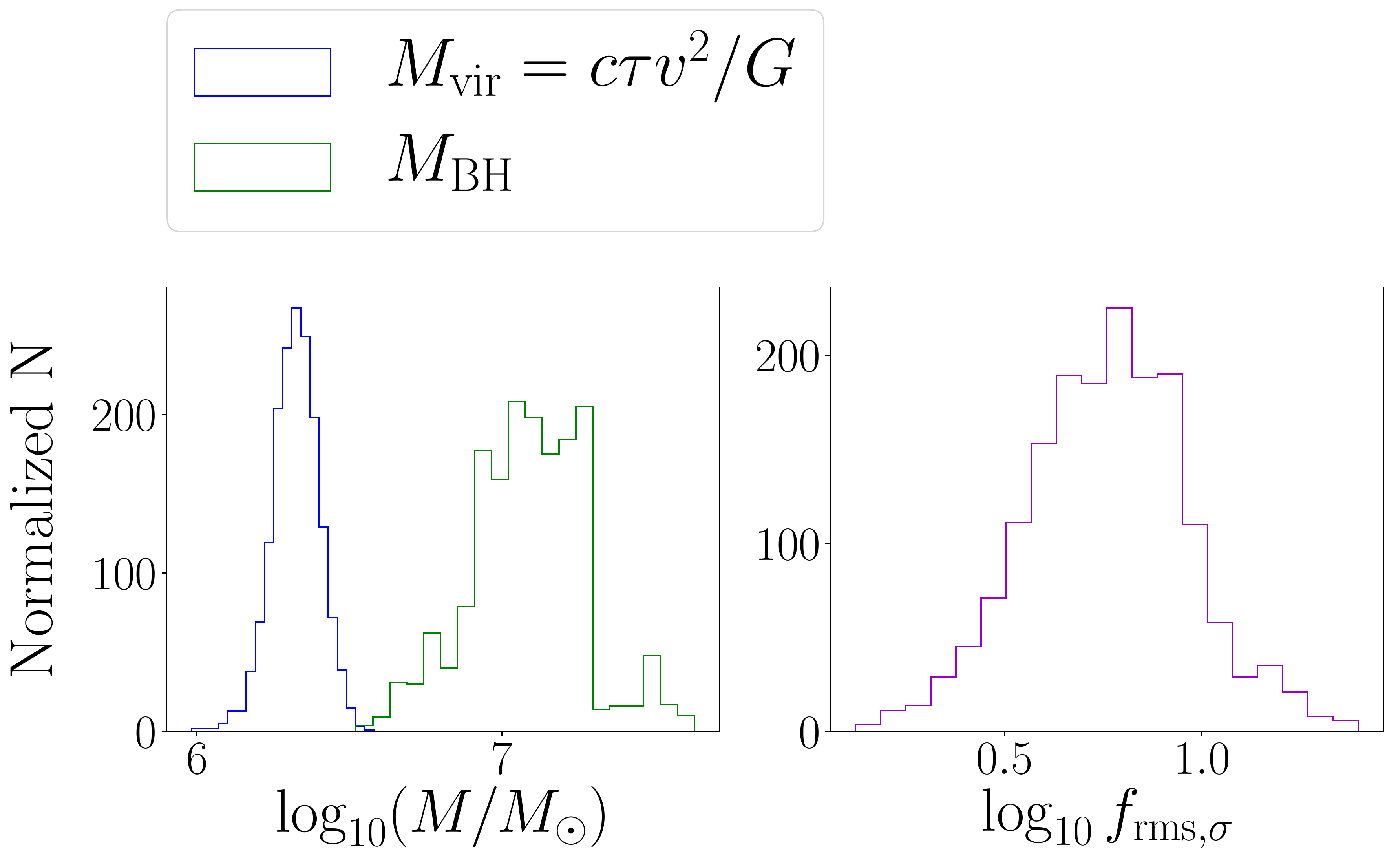}
    \caption{Taking random draws from the Gaussian distributions of the cross-correlation time lag and line widths (Figure \ref{fig:virial_product}), we calculate the virial product (shown in blue in the left panel) until the size of the virial product distribution is the same as that of the \textsc{caramel} \mbh\ posterior distribution function (shown in green in the left panel). The logarithmic virial coefficient of any given source in our sample is found by subtracting the logarithmic virial product distribution from the logarithmic \textsc{caramel} \mbh\ posterior distribution (i.e., dividing the original, nonlogarithmic distributions). The resulting distribution of logarithmic scale factor (\logfrmssigma) is shown on the right panel, which allows us to report errors on our measurement by quoting the 68\% confidence interval as define by the distribution. }
    \label{fig:calc_f}
\end{figure}

Then, we take random draws from the Gaussian distributions of the $\tau_{\rm{cen}}$ and $v$ measurements, and calculate the virial product, $M_{\rm{vir}}=c\tau v^2/ G$, until the number of draws is equal to the size of the \mbh\ posterior sample produced by \textsc{caramel}. Finally, we find a posterior distribution for the scale factor by dividing the \mbh\ distribution produced by \textsc{caramel} by the virial product distribution created above (see Figure \ref{fig:calc_f} below). From the posterior distribution produced, we report the median value and use a 68\% confidence interval for 1$\sigma$ uncertainties. Results for the individual scale factors of the LAMP 2016 sample are listed in Table~\ref{tab:table_individual_f}.

\begin{deluxetable*}{lcccc}
\tablecaption{Inferred Scale Factors}
\tablewidth{0pt}
\setlength{\tabcolsep}{10pt}
\tablehead{
\colhead{Galaxy} &
\colhead{$\log_{10}(f_{\rm{rms,}\sigma})$} &
\colhead{$\log_{10}(f_{\rm{rms, FWHM}})$} &
\colhead{$\log_{10}(f_{\rm{mean,}\sigma})$} &
\colhead{$\log_{10}(f_{\rm{mean, FWHM}})$}
}
\startdata
PG 2209+184 & $0.84^{+0.21}_{-0.20}$ & $0.08^{+0.21}_{-0.20}$ & $0.71^{+0.21}_{-0.19}$ & $-0.11^{+0.21}_{-0.20}$\\
\mcga & $1.21^{+0.41}_{-0.31}$ & $0.54^{+0.41}_{-0.31}$ & $1.08^{+0.41}_{-0.31}$ & $0.34^{+0.41}_{-0.31}$ \\
Mrk 1392 & $1.10^{+0.12}_{-0.14}$ & $0.32^{+0.13}_{-0.14}$ & $1.02^{+0.12}_{-0.14}$&$0.19^{+0.12}_{-0.14}$ \\
RBS 1303 & $0.05^{+0.27}_{-0.20}$ & $-0.23^{+0.25}_{-0.17}$ & $0.07^{+0.24}_{-0.16}$ & $-0.46^{+0.23}_{-0.16}$ \\
RBS 1917 & $0.82^{+0.32}_{-0.34}$ & $0.54^{+0.26}_{-0.32}$ & $0.54^{+0.27}_{-0.32}$ & $-0.08^{+0.50}_{0.32}$ \\
Mrk 841 & $0.62^{+0.50}_{-0.34}$& $-0.40^{+0.47}_{-0.38}$&$0.67^{+0.50}_{-0.36}$& $-0.38^{+0.50}_{-0.36}$ \\
RXJ 2044.0+2833 & $0.76^{+0.19}_{-0.19}$ & $0.02^{+0.18}_{-0.20}$& $0.66^{+0.18}_{-0.20}$& $-0.04^{+0.18}_{-0.20}$ \\
NPM1G+27.0587 & $0.98^{+0.51}_{-0.47}$ & $0.53^{+0.52}_{-0.46}$ & $1.01^{+0.50}_{-0.47}$ & $0.37^{+0.52}_{-0.46}$ \\
Mrk 1048 & $1.05^{+0.65}_{-0.57}$ & $0.33^{+0.64}_{-0.61}$ & $1.00^{+0.66}_{-0.57}$ & $0.16^{+0.66}_{-0.58}$ \\
\enddata
\tablecomments{Individual scale factors for the nine LAMP 2016 sources modeled by \citetalias{Villafa_a_2022}. Values were determined using our model \mbh\ estimates and corresponding line widths and cross correlation time lags found by \citet{u2021lick}. Individual scale factors of our extended sample can be found in their respective \textsc{caramel} papers: \citetalias{pancoast14b,Grier++17,2018ApJ...866...75W,2020ApJ...902...74W,bentz2021detailed}; and \citetalias{2022ApJ...934..168B}.}
\label{tab:table_individual_f}
\end{deluxetable*}

\subsubsection{Extended Sample}
We extend our sample (see Table \ref{tab: extended_sample}) by combining our results with prior \textsc{caramel} studies --- namely seven from LAMP 2011  \citepalias{2018ApJ...866...75W}, four from AGN10  \citepalias{Grier++17}, five from LAMP 2008 \citepalias{pancoast14b}, one from AGNSTORM \citepalias{2020ApJ...902...74W}, one from  \citetalias{bentz2021detailed}, and one from \citetalias{2022ApJ...934..168B}. 

The line widths used to compute the virial coefficient using the approach described above can be found in Table \ref{tab: line_widths}. All line widths correspond to those used in our previous \textsc{caramel} studies, with the exception of the four from the AGN 10 \citepalias{Grier++17} campaign. The line widths previously used did not have the narrow-line component removed. In order to remain consistent within our extended sample when searching for correlations with line-profile shape, we remeasured the line widths of these four points using the data from \citetalias{Grier++17}, in which the narrow-line contribution had been removed. To remeasure these line widths, we used the methods of \citet{u2021lick} and computed a Monte Carlo bootstrapping procedure for error analysis. 

The values of the individual AGN-specific virial coefficients are also found in their respective \textsc{caramel} papers and were determined in the same fashion as the LAMP 2016 sample described above. 

\begin{deluxetable*}{llcccc}
\setlength{\tabcolsep}{9pt}
\tablecaption{Extended Sample}
\tablehead{
\colhead{Campaign} & 
\colhead{Galaxy} & 
\colhead{Redshift } &
\colhead{$\log_{10}(\rm{M}_{\rm{BH}}/\rm{M}_{\odot})$}}
\startdata
\multirow{ 5}{*}{\begin{tabular}{l} Lick AGN Monitoring Project \\ \citep[LAMP 2008;][hereafter \citetalias{pancoast14b}]{pancoast14b}  \\ \end{tabular}} & Arp 151 & 0.02109 & $6.62^{+0.10}_{-0.13}$ \\
& Mrk 1310 & 0.01941 & $7.42^{+0.26}_{-0.27}$ \\
& NGC 5548 & 0.01718 & $7.51^{+0.23}_{-0.14}$\\
& NGC 6814 & 0.00521 & $6.42^{+0.24}_{-0.18}$\\
& SBS 1116+583A & 0.02787 & $6.99^{+0.32}_{-0.25}$\\ \hline
\multirow{ 4}{*}{\begin{tabular}{l} 2010 AGN monitoring campaign at MDM Observatory  \\ \citep[AGN10;][hereafter \citetalias{Grier++17}]{Grier++17} \\ \end{tabular}} & Mrk 335  & 0.0258 & $7.25^{+0.10}_{-0.10}$ \\
& Mrk 1501  & 0.0893 & $7.86^{+0.20}_{-0.17}$ \\
& 3C 120  & 0.0330 & $7.84^{+0.14}_{-0.19}$\\
& PG 2130+099  &0.0630 & $6.92^{+0.24}_{-0.23}$\\ \hline
\multirow{ 7}{*}{\begin{tabular}{l} Lick AGN Monitoring Project \\ \citep[LAMP 2011;][hereafter \citetalias{2018ApJ...866...75W}]{2018ApJ...866...75W} \\ \end{tabular}} & Mrk 50  & 0.0234 & $7.50^{+0.25}_{-0.18}$\\
& Mrk 141  & 0.0417 & $7.46^{+0.15}_{-0.21}$\\
& Mrk 279 &  0.0305 & $7.58^{+0.08}_{-0.08}$\\
& Mrk 1511  & 0.0339 &  $7.11^{+0.20}_{-0.17}$\\
& NGC 4593 & 0.0090 & $6.65^{+0.27}_{-0.15}$\\
& Zw 229-015  & $0.0279$ & $6.94^{+0.14}_{-0.14}$ \\ 
& PG 1310-108 & $0.0343$ & $6.48^{+0.21}_{-0.18}$ \\ \hline
\multirow{ 3}{*}{\begin{tabular}{l}  Space Telescope and Optical Reverberation Mapping Project \\ \citep[AGNSTORM;][hereafter \citetalias{2020ApJ...902...74W}]{2020ApJ...902...74W} \\ \end{tabular}} & & \\
& NGC $5548$ & $0.017175$ & $7.64^{+0.21}_{-0.18}$ \\ 
& & & \\ \hline
\multirow{ 3}{*}{\begin{tabular}{l}  AGN monitoring campaign at Las Cumbres Observatory \\ \citep[LCO;][hereafter  \citetalias{bentz2021detailed}]{bentz2021detailed} \\ \end{tabular}} & & \\
& NGC $3783$ & $0.097$ & $7.51 ^{+0.26}_{-0.13}$ \\
& & & \\ \hline
\multirow{ 9}{*}{\begin{tabular}{l} Lick AGN Monitoring Project \\ \citep[LAMP 2016;][hereafter \citetalias{Villafa_a_2022}]{Villafa_a_2022} \\\end{tabular}} & \pga  & $0.07000$ &  $7.53^{+0.19}_{-0.20}$  \\ 
& \rbsa & $0.06600$ & $7.04^{+0.23}_{-0.35}$  \\
& \mcga & $0.03235$ & $7.59^{+0.42}_{-0.28}$  \\
& NPM1G$+27.0587$ &  $0.06200$ & $7.64^{+0.40}_{-0.36}$\\
&\mrka  & $0.03614$ & $8.16^{+0.11}_{-0.13}$ \\
& \rbsb  & $0.04179$ & $6.79^{+0.19}_{-0.11}$\\
& Mrk $1048$ & $0.04314$ & $7.79^{+0.44}_{-0.48}$\\
& \rxja &  $0.05000$ & $7.09^{+0.17}_{-0.17}$ \\
& \mrke &  $0.03642$ &  $7.62^{+0.50}_{-0.30}$\\ \hline 
\multirow{2}{*}{\begin{tabular}{l}  AGN monitoring campaign at MDM Observatory \\ \citep[MDM;][hereafter \citetalias{2022ApJ...934..168B}]{2022ApJ...934..168B} \\ \end{tabular}} & & \\
& NGC $4151$ & $0.0033$ & $7.22^{+0.11}_{-0.10}$ \\
 & & & \\
\enddata
\tablecomments{Extended sample includes sources modeled by \citetalias{pancoast14b}, \citetalias{Grier++17}, \citetalias{2018ApJ...866...75W}, \citetalias{2020ApJ...902...74W}, \citetalias{bentz2021detailed}, and \citetalias{2022ApJ...934..168B}, in addition to the most recent sampled modeled by \citetalias{Villafa_a_2022}. Column 1 specifies the campaign from which data were collected and galaxy name is found in column 2. Columns 3 and 4 list the galaxy's redshift and \textsc{caramel} \mbh\ estimate, as defined by the 68\% confidence interval of the resultant posterior distribution function, respectively. } 
\label{tab: extended_sample}
\end{deluxetable*} 

\section{Results}\label{sec:RESULTS}
Using the individual AGN-specific virial coefficients determined for our extended sample, and enabled by our \textsc{caramel} forward modeling approach, we carry out a systematic investigation of correlations between $f$ and observables.

We use the \textsc{IDL} routine \texttt{linmix\_err} \citep{Kelly07} to perform a Bayesian linear regression in order to account for correlated measurement uncertainties. Doing so allows us to analyze the actual intrinsic correlation of any two parameters without worrying about a false increase due to correlated measurement uncertainties. This is especially important for our search for correlations with scale factor since individual scale factors are determined using our model \mbh\ measurements, and therefore uncertainties in the scale factor are connected to uncertainties in other model parameters.

To quantify the strength of any correlation, we compare the median fit slope to the $1\sigma$ uncertainty in the slope and determine our level of confidence using the following intervals we have defined in our previous study \citetalias{2018ApJ...866...75W}: we classify 0--$2\sigma$ as no evidence, 2--$3\sigma$ as marginal evidence, 3--$5\sigma$ as evidence, and $>5\sigma$ as conclusive evidence. 

Overall, we find the following correlations with at least marginal evidence as defined by our confidence intervals:
\begin{enumerate}
    \item \textbf{Black Hole Mass:}
    \begin{center}
      \subitem \logfmeansigma\ \ vs. $\log_{10}(M_{\rm{BH}}/M_{\odot})$; \\
      $\beta=0.51\pm0.15$, $\sigma_{\rm{int}}=0.22\pm0.05$, \\
    $3.4\sigma$ evidence \\ 
    \vspace{0.4cm}
     \subitem \logfrmssigma\ \ vs. $\log_{10}(M_{\rm{BH}}/M_{\odot})$ \\
      $\beta=0.47\pm0.17$, $\sigma_{\rm{int}}=0.25^{+0.06}_{-0.05}$, \\
    $2.8\sigma$ marginal evidence
    \end{center}
    \item \textbf{Opening Angle (BLR disk thickness):}
    \begin{center}
    \subitem \logfmeanfwhm\ vs. $\theta_o$; \\
    $\beta=-0.96^{+0.47}_{-0.43}$, $\sigma_{\rm{int}}=0.22^{+0.06}_{-0.05}$\\
    $2.1\sigma$ marginal evidence \\
    \vspace{0.4cm}
    \subitem \logfrmsfwhm\ vs. $\theta_o$; \\
    $\beta=-1.15^{+0.48}_{-0.46}$, $\sigma_{\rm{int}}=0.21^{+0.06}_{-0.05}$\\
    $2.4\sigma$ marginal evidence \\
    \vspace{0.4cm}
    \end{center}
    \item \textbf{Inclination Angle:}
    \begin{center}
    \subitem \logfmeanfwhm\ vs. $\theta_i$; \\
    $\beta=-1.45^{+0.53}_{-0.56}$, $\sigma_{\rm{int}}=0.17^{+0.05}_{-0.04}$\\
    $2.6\sigma$ marginal evidence \\
    \vspace{0.4cm}
    \subitem \logfrmssigma\ vs. $\theta_i$; \\
    $\beta=-1.61^{+0.66}_{-0.68}$, $\sigma_{\rm{int}}=0.22^{+0.06}_{-0.05}$\\
    $2.4\sigma$ marginal evidence \\
    \vspace{0.4cm}
    \subitem \logfmeansigma\ vs. $\theta_i$; \\
    $\beta=-1.37^{+0.66}_{-0.68}$, $\sigma_{\rm{int}}=0.23\pm0.05$\\
    $2.0\sigma$ marginal evidence 
    \end{center}
    \item \textbf{Line-Profile Shape:}
    \begin{center}
    \subitem (FWHM$/\sigma)_{\rm{rms}}$ vs. \logfrmssigma; \\
     $\beta=1.50^{+0.67}_{-0.71}$, $\sigma_{\rm{int}}=0.24^{+0.09}_{-0.08}$ \\
     $2.2\sigma$ marginal evidence
     \end{center}
\end{enumerate}

\subsection{Correlations between f \& AGN/BLR Parameters}
\label{sec: scale_factor_correlations}
\begin{figure*}
    \centering
    \includegraphics[height=7cm, keepaspectratio]{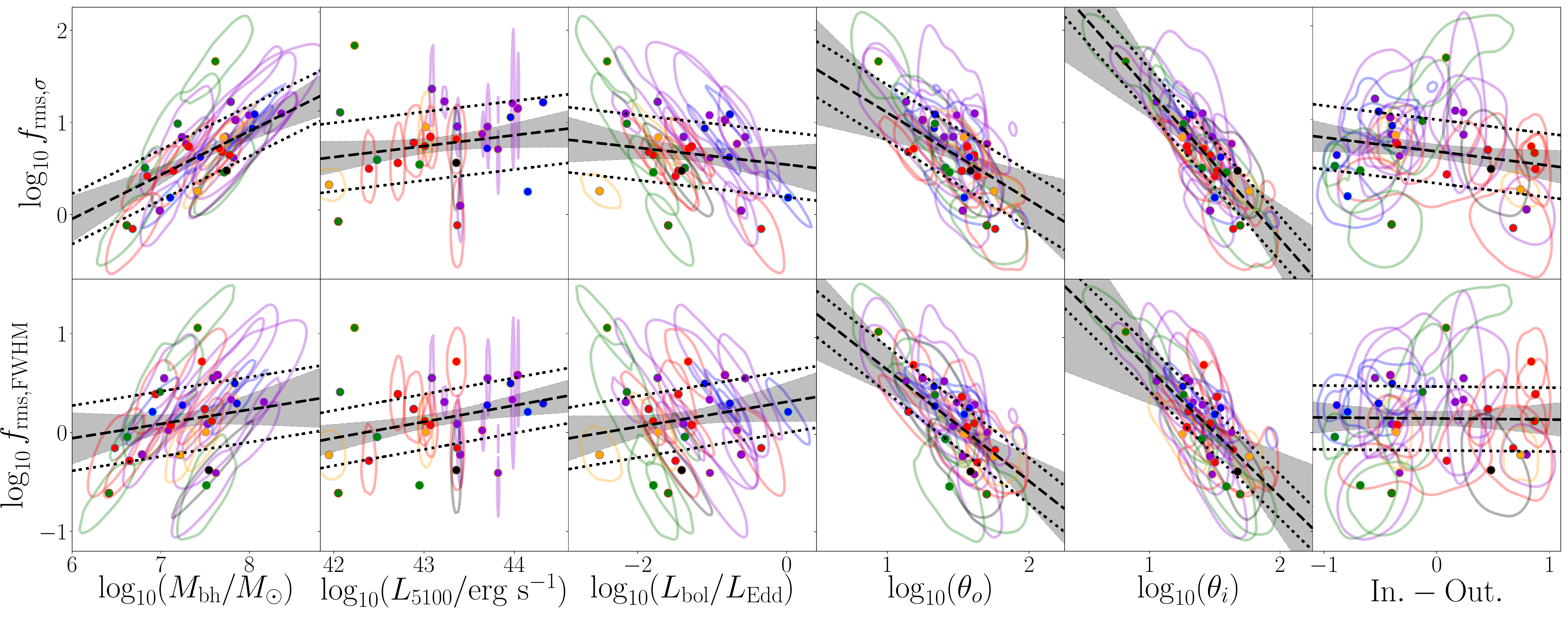}
    \caption{Correlations between the scale factor  $\log_{10}f_{\rm{rms},\sigma}$ (top) and $\log_{10}f_{\rm{rms, FWHM}}$ (bottom) with select AGNs and model parameters. From left to right: \mbh, optical luminosity, Eddington ratio, \Hb-emitting BLR opening angle (disk thickness), \Hb-emitting BLR inclination angle, and our ``inflow-outflow" parameter.  The colored dots and contours show the median and 68\% confidence regions of the 2D posterior PDFs for each AGN. 
    The dashed black lines and gray shaded regions give the median and 68\% confidence intervals of the linear regression. Dotted lines are offset above and below the dashed line by the median value of the intrinsic scatter. Purple points are for the AGNs from \citetalias{Villafa_a_2022}, red points are from \citetalias{2018ApJ...866...75W}, green points are from \citetalias{pancoast14b}, blue points are from \citetalias{Grier++17}, the black point is from \citetalias{2020ApJ...902...74W}, and the orange points are from \citetalias{bentz2021detailed} and \citetalias{2022ApJ...934..168B}.}
    \label{fig:f_rms_correlations}
\end{figure*}

\begin{figure*}
    \centering
    \includegraphics[height=7cm, keepaspectratio]{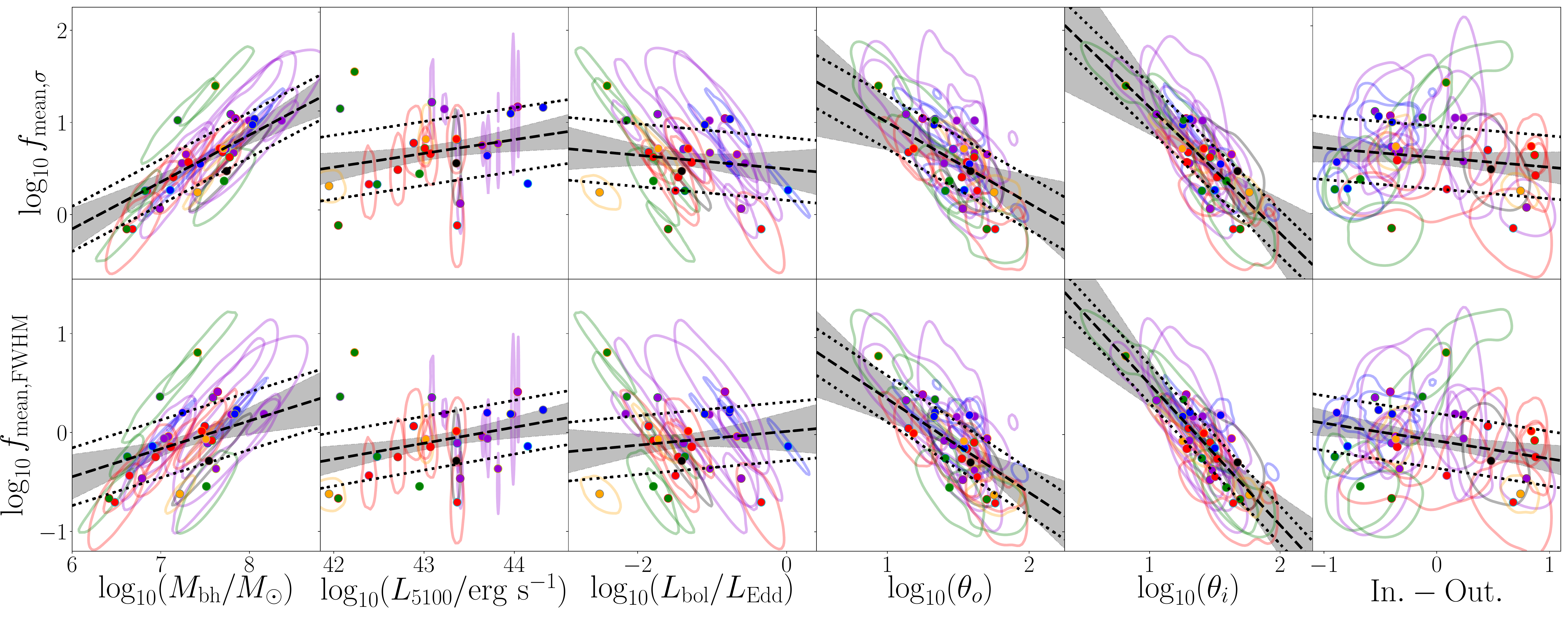}
    \caption{Correlations between the scale factor  $\log_{10}f_{\rm{mean,}\sigma}$ (top) and $\log_{10}f_{\rm{mean,FWHM}}$ (bottom) with select AGNs and model parameters. From left to right: \mbh, optical luminosity, Eddington ratio, \Hb-emitting BLR opening angle (disk thickness), \Hb-emitting BLR inclination angle, and our ``inflow-outflow" parameter. The colored dots and contours show the median and 68\% confidence regions of the 2D posterior PDFs for each AGN. 
    The dashed black lines and gray shaded regions give the median and 68\% confidence intervals of the linear regression. Dotted lines are offset above and below the dashed line by the median value of the intrinsic scatter. Purple points are for the AGNs from \citetalias{Villafa_a_2022}, red points are from \citetalias{2018ApJ...866...75W}, green points are from \citetalias{pancoast14b}, blue points are from \citetalias{Grier++17}, the black point is from \citetalias{2020ApJ...902...74W}, and the orange points are from \citetalias{bentz2021detailed} and \citetalias{2022ApJ...934..168B}.}
    \label{fig:f_mean_correlations}
\end{figure*}

\begin{deluxetable*}{c l c cccccc}
\tablecaption{Linear regression results for rms spectrum scale factors}
\tablewidth{0pt}
\setlength{\tabcolsep}{5pt}
\tablehead{
\colhead{$f$-type} &
\colhead{ } &
\colhead{$\log_{10}(\rm{M}_{\rm BH}/\rm{M}_{\odot})$ } &
\colhead{$\log_{10}(\rm{L}_{5100}/{\rm erg~s}^{-1})$ } &
\colhead{$\log_{10}(\rm{L}_{\rm bol}/\rm{L}_{\rm Edd})$ } &
\colhead{$\theta_o$ (${\rm deg.}$)} &
\colhead{$\theta_i$ (${\rm deg.}$)} &
\colhead{${\rm In.-Out.~param.}$ }
}
\startdata
& $\alpha$    & $-2.74^{+ 1.22} _{- 1.24}$ &  $-4.03^{+ 5.19} _{- 5.03}$ &   0.55$\pm0.21$ &  2.05$^{+ 0.81} _{- 0.90}$ &    2.95$^{+ 0.97} _{- 0.96}$ &   0.67$\pm0.08$ \\
${\rm rms},{\sigma}$ & $\beta$ & 0.47$\pm0.17$ &    0.11$\pm0.12$ &  $-0.09^{+ 0.14} _{- 0.15}$ &   $-0.95^{+ 0.60} _{- 0.54}$ &  $-1.61^{+ 0.66} _{- 0.68}$ &  $-0.16\pm0.14$  \\
 & $\sigma_{\rm int}$ & 0.25$^{+ 0.06} _{- 0.05}$ &   0.31$^{+ 0.09} _{- 0.08}$ &   0.32$^{+ 0.08} _{- 0.07}$ &   0.28$^{+ 0.07} _{- 0.05}$ &   0.22$^{+ 0.06} _{- 0.05}$ &   0.31$^{+ 0.08} _{- 0.07}$ \\
 \hline
 \multirow{ 3}{*}{\begin{tabular}{l}rms, \\ FWHM\end{tabular}}
 & $\alpha$     &   $-0.91^{+ 1.41} _{- 1.43}$ &  $-7.18^{+ 4.70} _{- 4.33}$ &  0.32$^{+ 0.19} _{- 0.20}$ & 1.82$^{+ 0.68} _{- 0.71}$ &  1.99$^{+ 0.99} _{- 1.16}$ &  0.15$\pm0.07$ \\
 &$\beta$  &   0.14$^{+ 0.20} _{- 0.19}$ &  0.17$^{+ 0.10} _{- 0.11}$ &  0.13$^{+ 0.13} _{- 0.14}$ &  $-1.15^{+ 0.48} _{- 0.46}$ &  $-1.31^{+ 0.80} _{- 0.69}$ &  $-0.01\pm0.14$ \\
&$\sigma_{\rm int}$ &   0.29$^{+ 0.08} _{- 0.07}$ &  0.24$\pm0.09$ &   0.27$^{+ 0.09} _{- 0.08}$ & 0.21$^{+ 0.06} _{- 0.05}$ &   0.20$^{+ 0.06} _{- 0.05}$ &  0.29$^{+ 0.09} _{- 0.08}$ \\
\enddata
\tablecomments{Linear regression results used to determine correlations between the scale factor $f$ and select AGNs and model parameters shown in Figure \ref{fig:f_rms_correlations}. The parameters $\alpha$ and $\beta$ represent the constant and slope of the linear regression, respectively. While $\sigma_{\rm{int}}$ represents the standard deviation of the intrinsic scatter. The corresponding relationship is therefore given by $\log_{10}f= \alpha + \beta\times \rm{parameter}+\mathcal{N}(0,\sigma_{\rm int})$.}
\label{tab:table_rms_correlations}
\end{deluxetable*}

\begin{deluxetable*}{c l c cccccc}
\tablecaption{Linear regression results for mean spectrum scale factors}
\tablewidth{0pt}
\setlength{\tabcolsep}{5pt}
\tablehead{
\colhead{$f$-type} &
\colhead{ } &
\colhead{$\log_{10}(\rm{M}_{\rm BH}/\rm{M}_{\odot})$ } &
\colhead{$\log_{10}(\rm{L}_{5100}/{\rm erg~s}^{-1})$ } &
\colhead{$\log_{10}(\rm{L}_{\rm bol}/\rm{L}_{\rm Edd})$ } &
\colhead{$\theta_o$ (${\rm deg.}$)} &
\colhead{$\theta_i$ (${\rm deg.}$)} &
\colhead{${\rm In.-Out.~param.}$ }
}
\startdata
 \multirow{ 3}{*}{\begin{tabular}{l}mean, \\ $\sigma$ \end{tabular}}  & $\alpha$   & $-3.11^{+ 1.06} _{- 1.10}$ &  $-5.47^{+ 4.87} _{- 4.61}$ & 0.51$^{+ 0.19} _{- 0.20}$ &     1.88$^{+ 0.75} _{- 0.88}$ &  2.53$^{+ 0.98} _{- 0.94}$&   0.60$\pm0.08$ \\
&$\beta$     &  0.51$\pm0.15$ &    0.14$\pm0.11$ &  $-0.07^{+ 0.13} _{- 0.14}$ &   $-0.88^{+ 0.59} _{- 0.50}$ &  $-1.37^{+ 0.66} _{- 0.68}$&  $-0.10\pm0.14$ \\
&$\sigma_{\rm {int}}$  &  0.22$\pm0.05$ &   0.29$^{+ 0.08} _{- 0.07}$ &   0.30$^{+ 0.08} _{- 0.06}$ &   0.26$^{+ 0.06} _{- 0.05}$ &  0.23$\pm0.05$  &    0.30$^{+ 0.08} _{- 0.07}$ \\ \hline
\multirow{ 3}{*}{\begin{tabular}{l}mean, \\ FWHM \end{tabular}}  &  $\alpha$ & $-1.99^{+ 1.26} _{- 1.28}$ &  $-7.13^{+ 4.37} _{- 4.04}$ &  0.01$\pm0.20$ &    1.33$^{+ 0.63} _{- 0.70}$ &   1.97$^{+ 0.81} _{- 0.75}$ &  $-0.08\pm 0.07$   \\
 & $\beta$   &  0.26$^{+ 0.18} _{- 0.17}$ &   0.16$^{+ 0.09} _{- 0.10}$ &  0.07$^{+ 0.13} _{- 0.14}$ &  $-0.96^{+ 0.47} _{- 0.43}$ &  $-1.45^{+ 0.53} _{- 0.56}$&  $-0.18\pm0.12$ \\
& $\sigma_{\rm {int}}$ &    0.26$\pm0.06$ &    0.23$^{+ 0.08} _{- 0.07}$ &   0.26$^{+ 0.08} _{- 0.07}$ &  0.22$^{+ 0.06} _{- 0.05}$ &   0.17$^{+ 0.05} _{- 0.04}$ &    0.24$^{+ 0.07} _{- 0.06}$  \\
\enddata
\tablecomments{Linear regression results used to determine correlations between the scale factor $f$ and select AGNs and model parameters shown in Figure \ref{fig:f_mean_correlations}. The parameters $\alpha$ and $\beta$ represent the constant and slope of the linear regression, respectively. While $\sigma_{\rm{int}}$ represents the standard deviation of the intrinsic scatter. The corresponding relationship is therefore given by $\log_{10}f= \alpha + \beta\times \rm{parameter} +\mathcal{N}(0,\sigma_{\rm int})$.}
\label{tab:table_mean_correlations}
\end{deluxetable*}

In an effort to uncover any regularities in the behavior of the BLR and gain a better understanding of the inner regions of AGNs, we investigate correlations between scale factor and AGN/BLR parameters determined by our forward-modelling approach. Overall, we find similar trends for both the rms and mean spectrum --- see Figures~\ref{fig:f_rms_correlations} and \ref{fig:f_mean_correlations} (respectively), and  Tables~\ref{tab:table_rms_correlations} and~\ref{tab:table_mean_correlations}, for their corresponding regression values. We reiterate that covariance between variables is taken into account in our analysis, in order to avoid spurious correlations. 

We find evidence ($3.4\sigma$) for a correlation between scale factor and \mbh\ when using the mean spectrum and line dispersion line width, i.e. \logfmeansigma\ \textbf{($\beta=0.51\pm0.15$)}, which had not been previously found by \citetalias{pancoast14b}, \citetalias{Grier++17}, or \citetalias{2018ApJ...866...75W}. Similarly, we find marginal evidence ($2.8 \sigma$) for a correlation between \logfrmssigma\ and \mbh\ \textbf{($\beta=0.47\pm0.17$)}. 
This correlation suggests that the BLR geometry and dynamics may be correlated with \mbh.

We also find marginal evidence for an anti-correlation with BLR opening angle, $\theta_o$, which is the \textsc{caramel} model parameter that represents the BLR disk thickness. When using FWHM  line-width measurements with both the mean \textbf{($\beta=-0.96^{+0.47}_{-0.43}$)} and rms \textbf{($\beta=-1.15^{+0.48}_{-0.46}$)} spectrum. Such a correlation with BLR disk thickness had also not been previously found in any previous \textsc{caramel} studies.

Finally, in agreement with previous results \citepalias[][]{pancoast14b,Grier++17, 2018ApJ...866...75W}, we find marginal evidence for an anti-correlation with BLR inclination angle and the virial coefficient, as measured when using the $\sigma$ line width with both the rms \textbf{($\beta=-1.61^{+0.66}_{-0.68}$)} and mean \textbf{($\beta=-1.37^{+0.67}_{-0.71}$)} spectra. Additionally, we also find marginal evidence for an anti-correlation when using the FWHM line width and the mean spectrum \textbf{($\beta=-1.45^{+0.53}_{-0.56}$)}.  This correlation was predicted by both \citet{collin06} and \citet{goad12}, and is expected for a disk-like BLR because an increase in BLR inclination angle would result in an increased observed line-of-sight velocity and therefore increased line-width measurement. Hence, in order to recover the same \mbh, a smaller scale factor would be required, producing an anti-correlation like the one that is apparent in our work. 

Before proceeding, it is important to note that although the correlations we have discovered with opening angle (BLR disk thickness) and inclination angle fall under our definition of marginally significant, they 
lack any real utility as BLR disk thickness is not an observable or a measurable quantity, and inclination-angle measurements using radio jets \citep[e.g.,][]{2005AJ....130.1418J, 2012ApJ...752...92A} are not possible for all cases. For these reasons, we now explore the existence of correlations between scale factor and a direct observable, line-profile shape --- that is, the ratio of the FWHM to the dispersion $\sigma_{\rm line}$, as such a correlation would provide an observational proxy for the virial coefficient, and thus a more reliable way to calibrate $f$.

\begin{deluxetable*}{lccccccc}
\tablecaption{Line Widths and Line Profile Shapes of Extended Sample}
\tablewidth{0pt}
\setlength{\tabcolsep}{7.5pt}
\tablehead{
\colhead{} &
\multicolumn{3}{c}{Rms } &
\multicolumn{3}{c}{Mean} \\
\colhead{Galaxy} & \colhead{FWHM} & \colhead{$\sigma_{\rm{line}}$} & \colhead{$\log_{10}(\rm{FWHM}/\sigma)$} & \colhead{FWHM} & \colhead{$\sigma_{\rm {line}}$} & \colhead{$\log_{10}(\rm{FWHM}/\sigma)$}
}
\startdata
Arp 151 \citepalias{pancoast11} &  $2458\pm82$\aref & $1295\pm37$\aref & $0.28\pm0.02$  & $3076\pm 39$\aref &  $1726\pm17$\aref & $0.25\pm0.007$\\
Mrk 1310 \citepalias{pancoast11} &  $1823\pm157$\aref & $921\pm135$\aref &$0.30\pm0.07$ &  $2425\pm19$\aref & $1229\pm12$\aref & $0.29\pm0.005$\\
NGC 5548 \citepalias{pancoast11} &  $12539\pm1927$\aref & $3900\pm266$\aref &$0.51^{+0.08}_{-0.07}$ & $12402\pm111$\aref & $4354\pm25$\aref & $0.45\pm0.004$\\
NGC 6814 \citepalias{pancoast11} & $2945\pm283$\aref & $1697\pm224$\aref & $0.24\pm0.07$ &  $3129\pm14$\aref & $1744\pm12$\aref & $0.25\pm0.003$\\
SBS 1116+583A \citepalias{pancoast11} & \nodata 
& \nodata & \nodata & $3135\pm36$\aref &  $1460\pm23$\aref &  $0.33\pm0.01$\\ 
Mrk 335 \citepalias{Grier++17} & $1853\pm79$\bref & $1239\pm78$\bref & $0.17\pm0.03$ & $2018\pm1$\bref & $1354\pm34$\bref & $0.17\pm0.01$\\
Mrk 1501 \citepalias{Grier++17} & $3476\pm214$\bref & $1401\pm48$\bref & $0.40\pm0.03$ & $3780\pm25$\bref & $1486\pm48$\bref & $0.41\pm0.01$\\
3C 120 \citepalias{Grier++17} & $2035\pm97$\bref & $1218\pm47$\bref & $0.22\pm0.03$ & $2893\pm22$\bref & $1175\pm26$\bref & $0.39\pm0.01$\\
PG 2130+099 \citepalias{Grier++17} & $1409\pm143$\bref & $1459\pm93$\bref & $-0.02\pm0.05$ & $2107\pm32$\bref & $1321\pm11$\bref & $0.20\pm0.01$\\
Mrk 50 \citepalias{2018ApJ...866...75W} &  $3355\pm128$\cref & $2020\pm103$\cref & $0.22\pm0.03$ &$4101\pm56$\cref &  $2024\pm31$\cref &  $0.31\pm0.01$\\
Mrk 141  \citepalias{2018ApJ...866...75W} & \nodata & \nodata & \nodata & $5129\pm45$\cref & $2280\pm21$\cref   &  $0.35\pm 0.01$\\
Mrk 279 \citepalias{2018ApJ...866...75W} &  $3306\pm338$\cref & $1778\pm71$\cref & $0.27\pm0.05$ &  $4099\pm43$\cref & $1821\pm13$\cref & $0.35\pm0.01$\\
Mrk 1511 \citepalias{2018ApJ...866...75W} &  $3236\pm65$\cref & $1506\pm42$\cref & $0.33^{+0.02}_{-0.01}$ &  $4154\pm28$\cref & $1828\pm12$\cref & $0.36\pm0.004$\\
NGC 4593 \citepalias{2018ApJ...866...75W} & $3597\pm72$\cref & $1601\pm40$\cref & $0.35\pm0.01$ & $4264\pm41$\cref & $1925\pm38$\cref &  $0.35\pm0.01$\\
Zw 229-015 \citepalias{2018ApJ...866...75W} & $1789\pm93$\cref & $1609\pm109$\cref & $0.05\pm0.04$& $3705\pm203$\cref & $1747\pm56$\cref &  $0.33\pm0.03$\\
PG 1310-108 \citepalias{2018ApJ...866...75W} & \nodata & \nodata & \nodata & $3422\pm21$\cref & $1823\pm20$\cref & $0.27\pm0.01$\\
NGC 5548 \citepalias{2020ApJ...902...74W} & $10861\pm739$\dref & $4115\pm513$\dref & $0.42^{+0.06}_{-0.07}$ & $9612\pm427$\dref & $3983\pm150$\dref & $0.38^{+0.02}_{-0.03}$\\
NGC 3783 \citepalias{bentz2021detailed} & $4278\pm676$\eref &  $1619\pm137$\eref &  $0.42^{+0.08}_{-0.07}$ &  $4486\pm35$\eref & $1825\pm19$\eref & $0.39\pm0.01$\\
PG 2209+184 \citepalias{Villafa_a_2022}& $3247\pm88$\fref & $1353\pm64$\fref & $0.38\pm0.02$  &  
$4045\pm34$\fref & 1573 $\pm40$\fref & $0.41\pm0.01$\\
\mcga \citepalias{Villafa_a_2022} & $2120\pm39$\fref & $977\pm29$\fref & 
$0.34^{+0.01}_{-0.02}$ & $2658\pm57$\fref & $1141\pm39$\fref & $0.37\pm 0.02$ \\
Mrk 1392 \citepalias{Villafa_a_2022} & $3690\pm138$\fref & $1501\pm38$\fref &  $0.39\pm0.02$ & $4267\pm25$\fref & $1635\pm13$\fref & $0.417\pm0.004$ \\
RBS 1303 \citepalias{Villafa_a_2022} &$1738\pm113$\fref & $1292\pm156$\fref &$0.13\pm0.06$ & $2286\pm21$\fref & $1243\pm26$\fref & $0.26\pm0.01$ \\
RBS 1917 \citepalias{Villafa_a_2022} & $1653\pm287$\fref & $851\pm154$\fref & $0.14^{+0.07}_{-0.09}$ & $2399\pm11$\fref & $1180\pm50$\fref & $0.31\pm0.02$\\
Mrk 841 \citepalias{Villafa_a_2022} &$7452\pm660$\fref & $2278\pm96$\fref  & $0.51\pm0.04$ & $7073\pm311$\fref & $2139\pm55$\fref & $0.52\pm0.02$\\
\rxja $~$\citepalias{Villafa_a_2022} & $2047\pm72$\fref & $870\pm50$\fref & $0.37\pm0.03$ &$2196\pm~31$\fref & $989\pm32$\fref &  $0.35^{+0.01}_{-0.02}$ \\
NPM1G+27.0587 \citepalias{Villafa_a_2022} &  $2893\pm177$\fref & $1735\pm136$\fref & $0.22^{+0.04}_{-0.05}$ & $3501\pm28$\fref & $1683\pm42$\fref & $0.32\pm0.01$\\
Mrk 1048 \citepalias{Villafa_a_2022} & $4042\pm406$\fref & $1726\pm76$\fref & $0.37\pm0.05$  & $4830\pm80$\fref & $1840\pm58$\fref & $0.42\pm0.02$ \\
NGC 4151 \citepalias{2022ApJ...934..168B} & $4711\pm750$\gref & $2680\pm64$\gref & $0.25^{+0.08}_{-0.06}$ & $7382\pm279$\gref & $2724\pm17$\gref &  $0.43\pm0.02$\\ 
\enddata
\tablecomments{All line widths are given in km s$^{-1}$. A line profile shape of $\log_{10}(\rm{FWHM}/\sigma)=0.371$ corresponds to a Gaussian profile, while $\log_{10}(\rm{FWHM}/\sigma) < 0.371$ corresponds to a Lorentz profile and $\log_{10}(\rm{FWHM}/\sigma) > 0.371$ corresponds to a flat topped profile. References for line widths are as follows: (a) \citet{2012ApJ...747...30P}, (b) This work --- the measurements used in previous \textsc{caramel} studies originated from \citet{2012ApJ...755...60G}, which did not remove the narrow line contribution. Thus, we remeasured using the data and spectral decompositions used by \citetalias{Grier++17}, in order to ensure these line width measurements were consistent with the rest of the sample, i.e. with the narrow line contribution removed. (c) \citet{2015ApJS..217...26B}, (d) \citet{2017ApJ...837..131P}, (e) \citet{2021ApJ...906...50B}, (f) \citet{u2021lick}, (g) \citet{2006ApJ...651..775B}.}
\label{tab: line_widths}
\end{deluxetable*}

\subsection{Line-Profile Shape as an Observational Proxy}
\label{sec: scalefactor_lineprofileshape}
\begin{figure*}
\centering
\includegraphics[height=8.5cm,keepaspectratio]{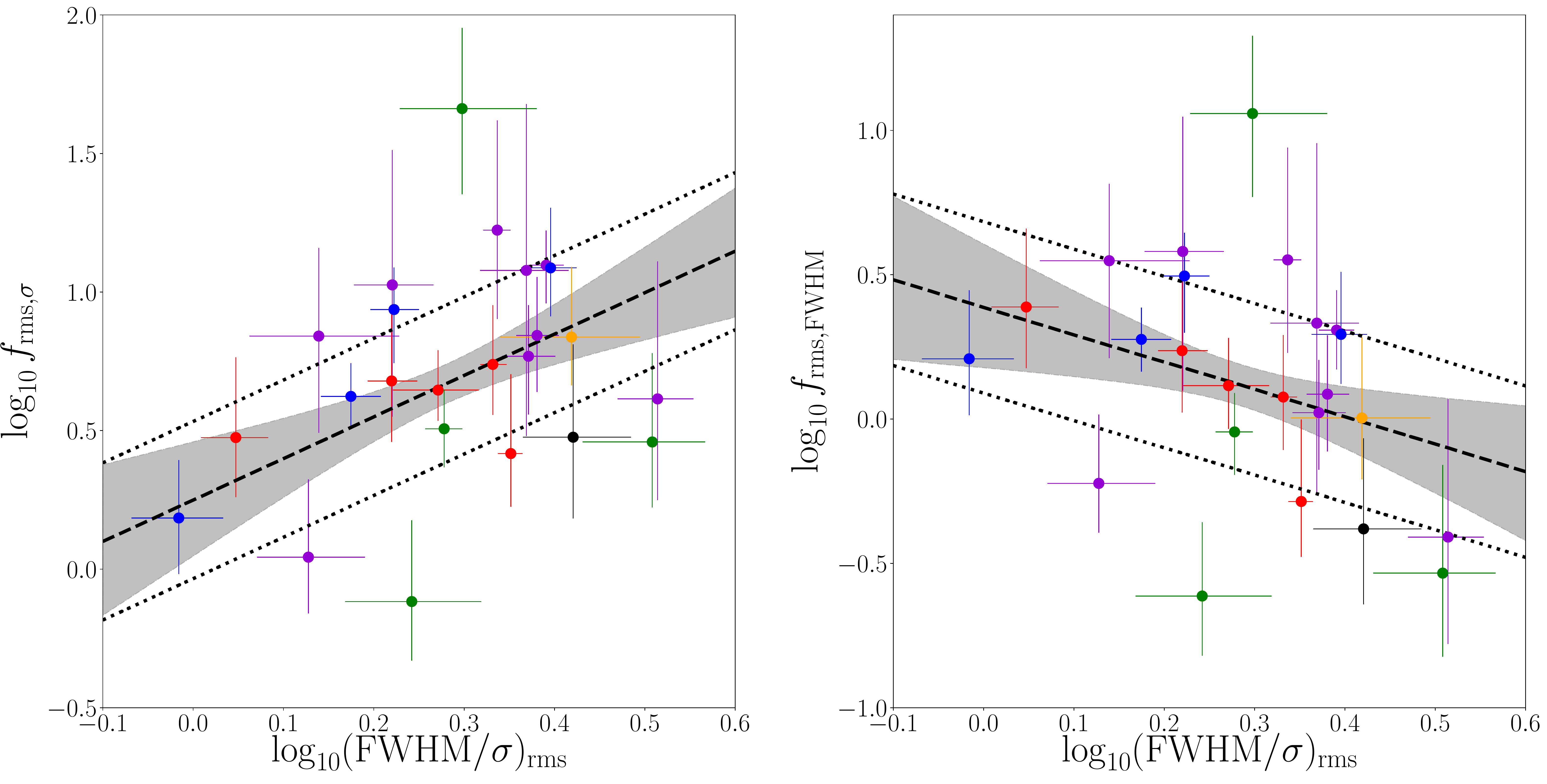}
\caption{Correlations between rms line-profile shape and scale factor determined using line dispersion (left) and FWHM (right). The dashed black lines and gray shaded regions give the median and 68\% confidence intervals of the linear regression. Dotted lines are offset above and below the dashed line by the median value of the intrinsic scatter. Purple points are for the AGNs from \citetalias{Villafa_a_2022}, red points are from \citetalias{2018ApJ...866...75W}, green points are from \citetalias{pancoast14b}, blue points are from \citetalias{Grier++17}, the black point is from \citetalias{2020ApJ...902...74W}, and the two orange points are from \citetalias{bentz2021detailed} and \citetalias{2022ApJ...934..168B}.}
\label{rms_line_profile_shape_withf}
\end{figure*}

\begin{figure*}
\centering
\includegraphics[height=8.5cm,keepaspectratio]{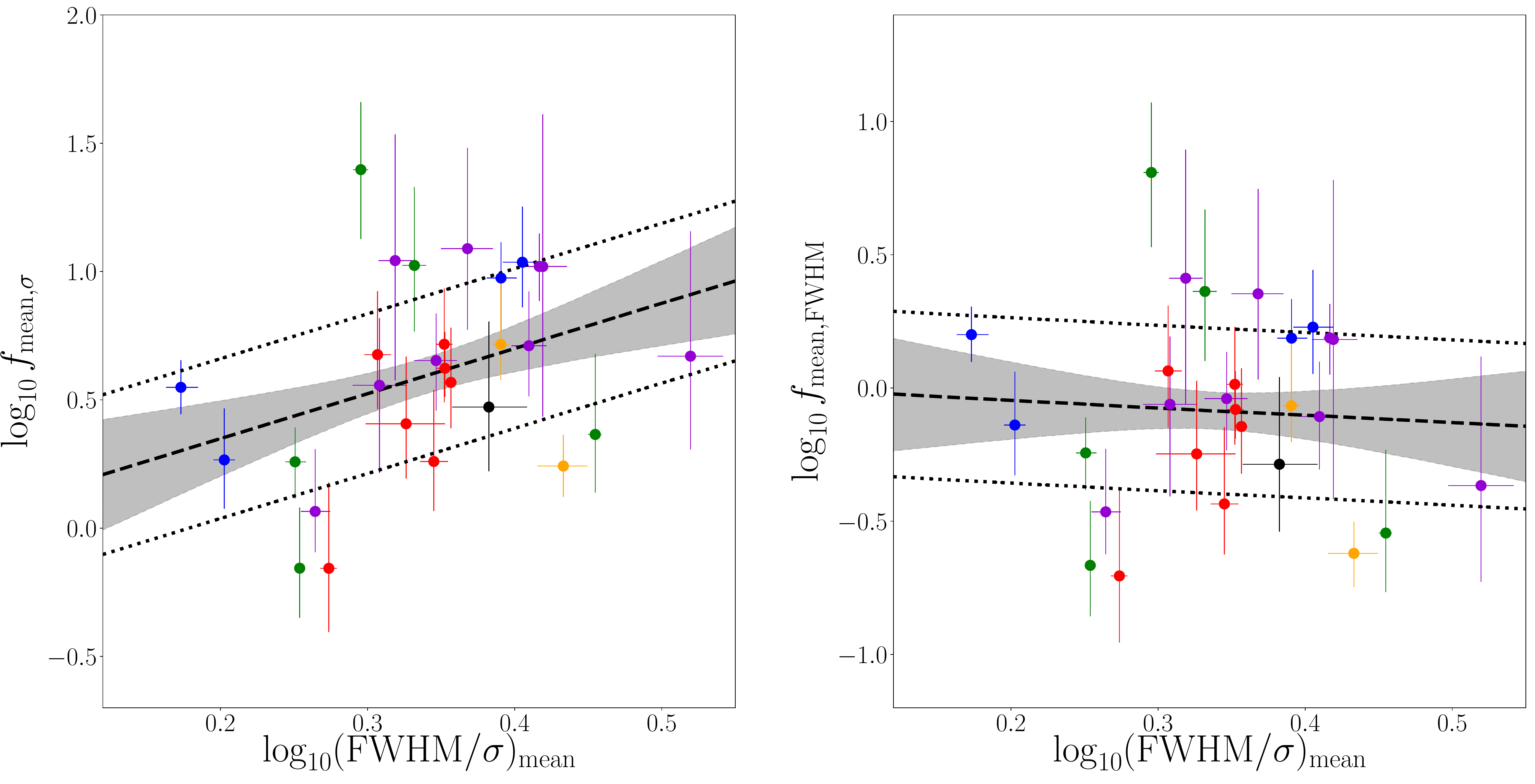}
\caption{Correlations between mean line-profile shape and scale factor determined using line dispersion (left) and FWHM (right). The dashed black lines and gray shaded regions give the median and 68\% confidence intervals of the linear regression. Dotted lines are offset above and below the dashed line by the median value of the intrinsic scatter. Purple points are for the AGNs from \citetalias{Villafa_a_2022}, red points are from \citetalias{2018ApJ...866...75W}, green points are from \citetalias{pancoast14b}, blue points are from \citetalias{Grier++17}, the black point is from \citetalias{2020ApJ...902...74W}, and the two orange points are from \citetalias{bentz2021detailed} and \citetalias{2022ApJ...934..168B}.}
\label{mean_line_profile_shape_withf}
\end{figure*}

We search for correlations with scale factor and line-profile shape using both the rms and mean spectrum (see Figures \ref{rms_line_profile_shape_withf} and \ref{mean_line_profile_shape_withf}, respectively), where we have used only the shape of the \Hb broad emission line by itself (i.e., we have isolated the broad emission from the narrow emission component). Line widths and line-profile shapes used for our extended sample are listed in Table \ref{tab: line_widths}. 

We find marginal evidence ($2.2\sigma$) for a correlation between \logfrmssigma\ and line-profile shape, when using the rms spectrum ($\beta=1.50^{+0.67}_{-0.71}$). When using the mean spectrum, however, the correlation falls short of being considered marginal evidence and is quantified by $1.9\sigma$ (see the left-most panels of Figure \ref{rms_line_profile_shape_withf} and \ref{mean_line_profile_shape_withf}, respectively). We do not find any evidence for a correlation with the virial coefficient when a FWHM line width is used, in either the rms or mean spectrum (see the right-most panels of Figures \ref{rms_line_profile_shape_withf} and \ref{mean_line_profile_shape_withf}).

Although stronger evidence is needed to recommend the widespread use of this relationship, this result is promising; further investigations with increased sample size in our dynamic modeling may help elucidate the correlation we have found.
\begin{deluxetable}{c l c cccccc}
\tablecaption{Linear regression results for line profile shape \\ vs. scale factor}
\setlength{\tabcolsep}{6pt}
\tablehead{
\colhead{Line Profile Shape} &
\colhead{ } &
\colhead{$\log_{10}f_{\sigma}$} &
\colhead{$\log_{10}f_{\rm{FWHM}}$}
}
\startdata
\multirow{ 3}{*}{\begin{tabular}{l} $\log_{10}\Big(\frac{\rm{FWHM}}{\sigma}\Big)_{\rm{mean}}$ \\ \end{tabular}}  & $\alpha$ &
 $-0.001\pm0.32$ & 0.01$^{+ 0.31} _{- 0.32}$ \\
 & $\beta$ & 1.76$^{+ 0.94} _{- 0.93}$ & $-0.28^{+ 0.93} _{- 0.92}$ \\
 & $\sigma_{\rm int}$ & 0.28$^{+ 0.08} _{- 0.06}$ & 0.28$^{+ 0.08} _{- 0.06}$ \\ \hline
 \multirow{ 3}{*}{\begin{tabular}{l} $\log_{10}\Big(\frac{\rm{FWHM}}{\sigma}\Big)_{\rm{rms}}$ \\ \end{tabular}}  & $\alpha$ & 0.25$^{+ 0.21} _{-0.20}$  & 0.39$^{+ 0.22} _{- 0.21}$ \\
 & $\beta$  & 1.50$^{+ 0.67} _{- 0.71}$ & $-0.95^{+ 0.69} _{- 0.73}$ \\
 & $\sigma_{\rm int}$  &  0.24$^{+ 0.09} _{-0.08}$  &   0.26$^{+ 0.09} _{- 0.07}$ \\
\enddata
\tablecomments{Linear regression results for line profile shape vs. scale factor. The parameters $\alpha$ and $\beta$ represent the constant and slope of the regression, respectively, while $\sigma_{\rm{int}}$ represents the standard deviation of the intrinsic scatter. The corresponding relationship is therefore given by $\log_{10}(\textit{f})= \alpha + \beta  \log_{10}(\rm{FWHM/}\sigma) +\mathcal{N}(0,\sigma_{\rm int})$.}
\label{tab:mean_lineprofile_f_linear_regression}
\end{deluxetable}

\section{The Role of BLR Geometry and Kinematics on Line-Profile Shape} \label{sec: toymodels}
From the correlations found in our work, we focus on the correlation found with $\log_{10}(\rm{FWHM}/\sigma)$, which has significant potential to improve the way the virial coefficient is calibrated. The relationship has an intrinsic scatter of similar magnitude to that of the local \mbh--$\sigma_*$ relation (see Table \ref{tab:mean_lineprofile_f_linear_regression}), which suggests another intrinsic relation of AGNs, and further validates the idea of using the line-profile shape of broad emission lines as a tracer for the inner regions of AGNs \citep{collin06}. In an attempt to gain a better understanding, we employ \textsc{caramel} models to test how BLR geometry and kinematics affects line-profile shape.

In particular, we aim to understand the line profiles with $\log_{10}(\rm{FWHM}/\sigma)_{\rm{mean}} \approx 0.1$--0.2. While line profiles with $\log_{10}(\rm{FWHM}/\sigma) = 0.37$ are best described by a Gaussian and are due to rotational Doppler broadening, Lorentz profiles (e.g.,  $\log_{10}(\rm{FWHM}/\sigma) < 0.37$) are thought to be a result of turbulent and/or inflow/outflow motions \citep{2013A&A...549A.100K}. In the following subsections, we investigate the effect of BLR size, disk thickness, inflow/outflow motion, and turbulent motion on \Hb broad line-profile shapes.

\subsection{BLR Size}
We begin by testing the effect of BLR size, since the extended wings in a Lorentz profile are due to high-velocity gas near the black hole. Thus, assuming Keplerian orbits, we expect narrower line-profile shapes to correspond to smaller BLR radii. We manipulate the \textsc{caramel} model parameters associated with BLR radius, $\mu$ and $F$, while keeping all other parameters constant. The values chosen for other relevant geometry and kinematics are as follows: $\theta_i=20^{\circ}$, $\theta_o=20^{\circ}$, 
$\beta=1.0$, $\log_{10}($\mbh$/M_{\odot})=7.5$, 
$f_{\rm{flow}}=0.5$, $f_{\rm{ellip}}=1.0$, $\theta_e=90^{\circ}$, $\sigma_{\rm{turb}}=0.001$. We choose the parameters to reflect particles with bound circular orbits (no inflow/outflow motion) and minimal contribution from macroturbulent velocities.

\begin{figure}
\centering
\includegraphics[height=6.0cm,keepaspectratio]{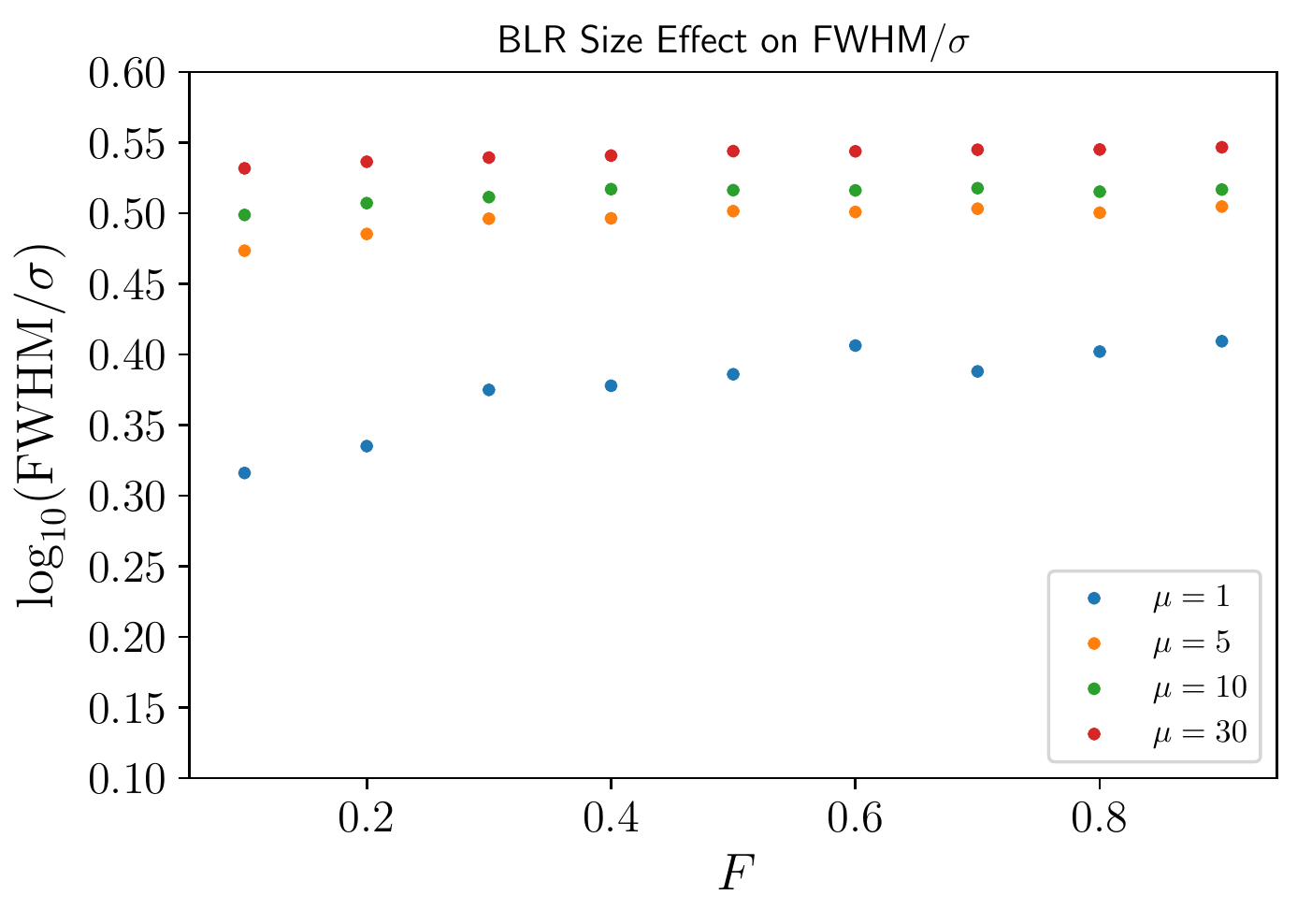}
\caption{We investigate the role of BLR radius in line profile shape using \textsc{caramel} models by varying the parameters $\mu$ and $F$, and holding all other model parameters constant. As described in the text, the parameter $\mu$ defines the mean BLR radius and the parameter $F$ defines the minimum radius in units of $\mu$. Different mean BLR radii, $\mu$, are depicted in different colors: 1 light-day is shown in blue, 5 light-days is shown in orange, 10 light-days is shown in green, and 30 light-days is shown in red. As expected, smaller values of $\log_{10}(\rm{FWHM}/\sigma)_{\rm{mean}}$ on the \textit{y}-axis, are seen with decreasing $\mu$. Additionally, within the four different mean radii, $\mu$, a slight decrease is seen for a decrease in minimum radius, as depicted by decreasing values of $F$ shown on the \textit{x}-axis.} 
\label{fig: blrsize_lineprofile}
\end{figure}

As shown in Figure \ref{fig: blrsize_lineprofile}, we find a smaller BLR size produces a smaller value of $\log_{10}(\rm{FWHM}/\sigma)$, as expected. However, we do not find any line-profile shapes in the region of  special interest, $\log_{10}(\rm{FWHM}/\sigma)_{\rm{mean}} \approx 0.1$--0.2, which suggests that bound circular orbits cannot produce these particular broad-line-profile shapes. Given the result that smaller BLR sizes produce smaller values of $\log_{10}(\rm{FWHM}/\sigma)$, and our ultimate goal of investigating what BLR geometry and kinematics produce smaller line-profile shapes, the remaining of our \textsc{caramel} model tests will focus solely on BLR sizes with mean radius, $\mu = 1$ and minimum radius within the range $F=0$--0.3.

\subsection{BLR Disk Thickness}
Next we test whether BLR disk thickness plays a role in determining the \Hb broad-line-profile shape. This idea stems from \citet{1981ARA&A..19..137P}, who found that the geometric height of an accretion disk is proportional to the ratio of turbulent velocity to rotational velocity of the disk. And although this notion applies to accretion disks, disk-outflow models which suggest that the BLR and the obscuring torus are closely connected, possibly forming one continuous structure that feeds/flows from the central accretion disk \citep[e.g.,][]{1992ApJ...385..460E, 1994ApJ...434..446K,1996VA.....40..133K, 2011A&A...536A..78K, 2015toru.conf..O08K}, qualify the application to a BLR disk. Thus, as suggested by \citet{2011BaltA..20..400K}, BLR lines with smaller values of $\log_{10}(\rm{FWHM}/\sigma)$ must have more of a spherical structure. 

\begin{figure}
\centering
\includegraphics[height=6.0cm,keepaspectratio]{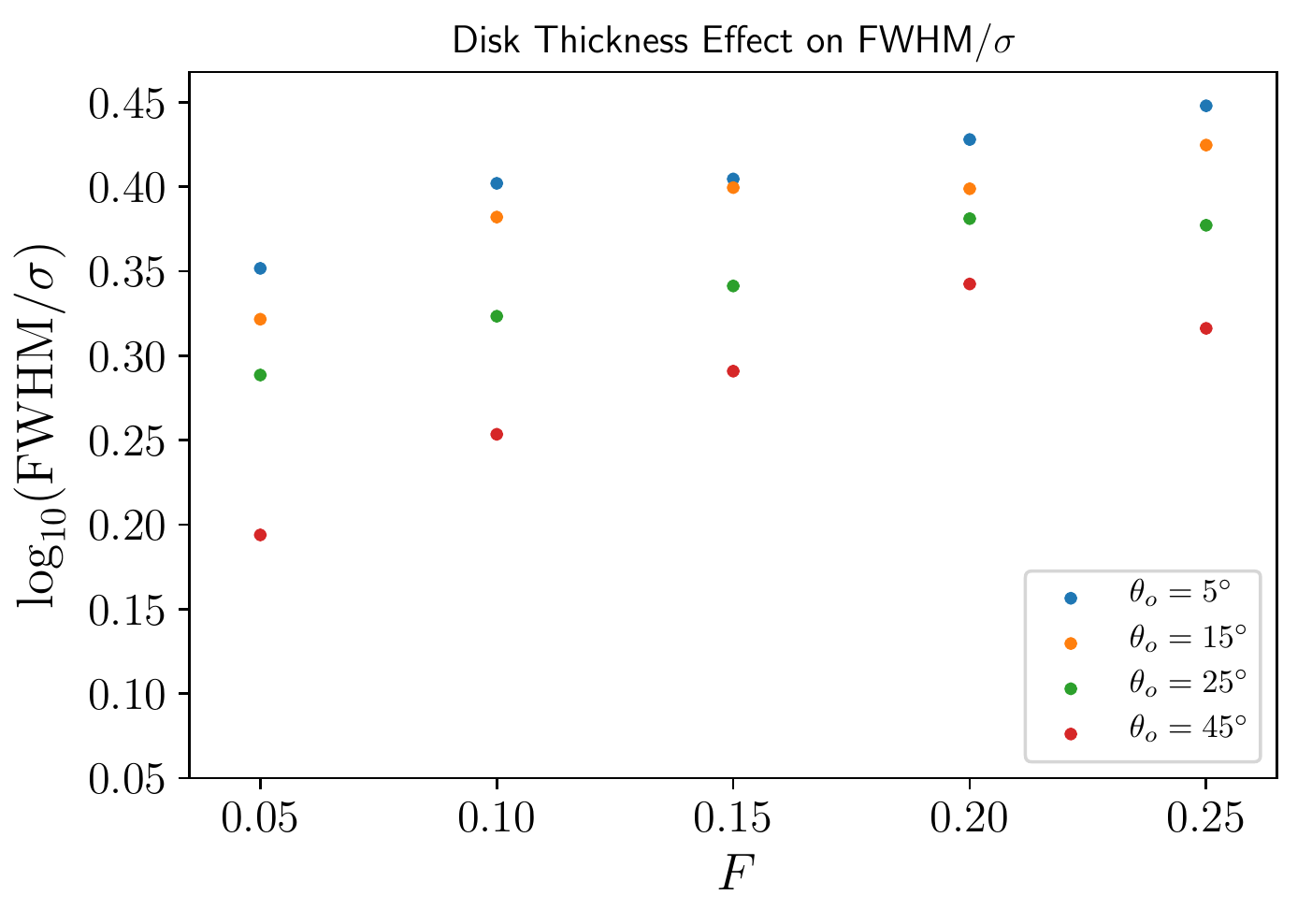}
\caption{We investigate the role of BLR disk thickness in line profile shape using \textsc{caramel} toy models by varying the parameter $\theta_o$ and minimum radius $F$, while holding all other model parameters constant. A mean radius of $\mu=1$ light-day is used, while minimum radius as defined by $F$ is varied using values $F=0-0.3$, as depicted by the \textit{x}-axis. Different BLR disk thickness/opening angles, $\theta_o$, are depicted in different colors: $\theta_o=5^{\circ}$ is shown in blue, $\theta_o=15^{\circ}$ is shown in orange, $\theta_o=25^{\circ}$ is shown in green, $\theta_o=45^{\circ}$ is shown in red. An opening angle of $\theta_o=45^{\circ}$, shown in red, corresponds to a spherical structure and produces broad lines with smaller values of $\log_{10}(\rm{FWHM}/\sigma)_{\rm{mean}}$, as expected.} 
\label{fig: blropeningangle_lineprofile}
\end{figure}

Using our \textsc{caramel} models, we vary $\theta_o$ and $F$, and keep all other parameters set to the following values: $\theta_i=25^{\circ}$, 
$\beta=1.0$, $\mu=1$, $\log_{10}($\mbh$/M_{\odot})=7.5$, 
$f_{\rm{flow}}=0.5$, $f_{\rm{ellip}}=1.0$, $\theta_{e}=90^{\circ}$, $\sigma_{\rm{turb}}=0.001$. Again, this configuration was selected in order to reflect particles on bound circular orbits. As expected, larger opening angles $\theta_o$ (i.e., thicker BLR disks) produce broad lines with smaller values of $\log_{10}(\rm{FWHM}/\sigma)_{\rm{mean}}$ (see Figure \ref{fig: blropeningangle_lineprofile}). The spherical BLR disk represented by $\theta_o=45^{\circ}$ even begins to have a line-profile shape defined by $\log_{10}($FWHM$/\sigma)\approx 0.2$, with bound circular orbits (without inflow/outflow and/or turbulent motion).

\begin{figure*}[ht]
\gridline{\fig{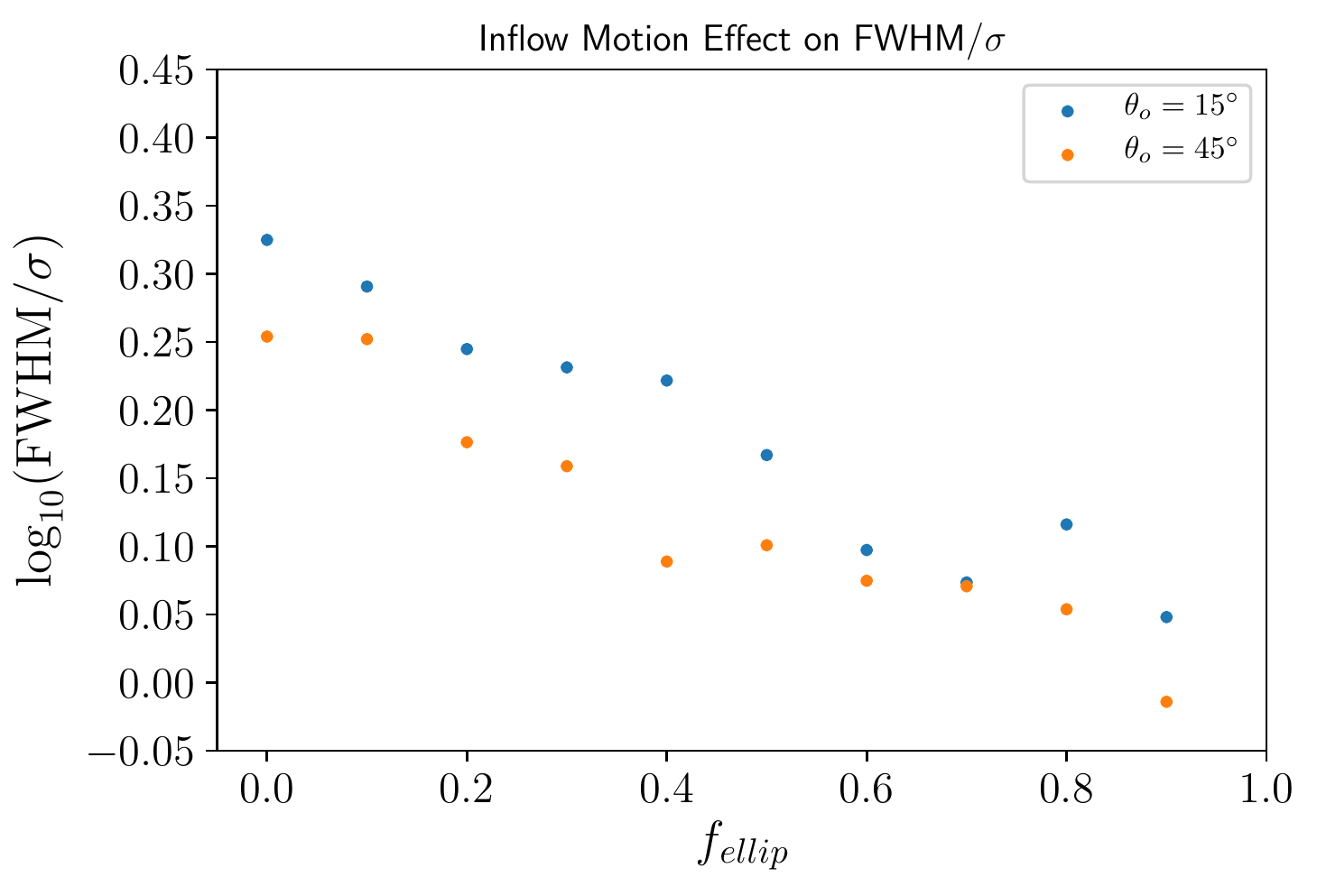}{0.5\textwidth}{}
          \fig{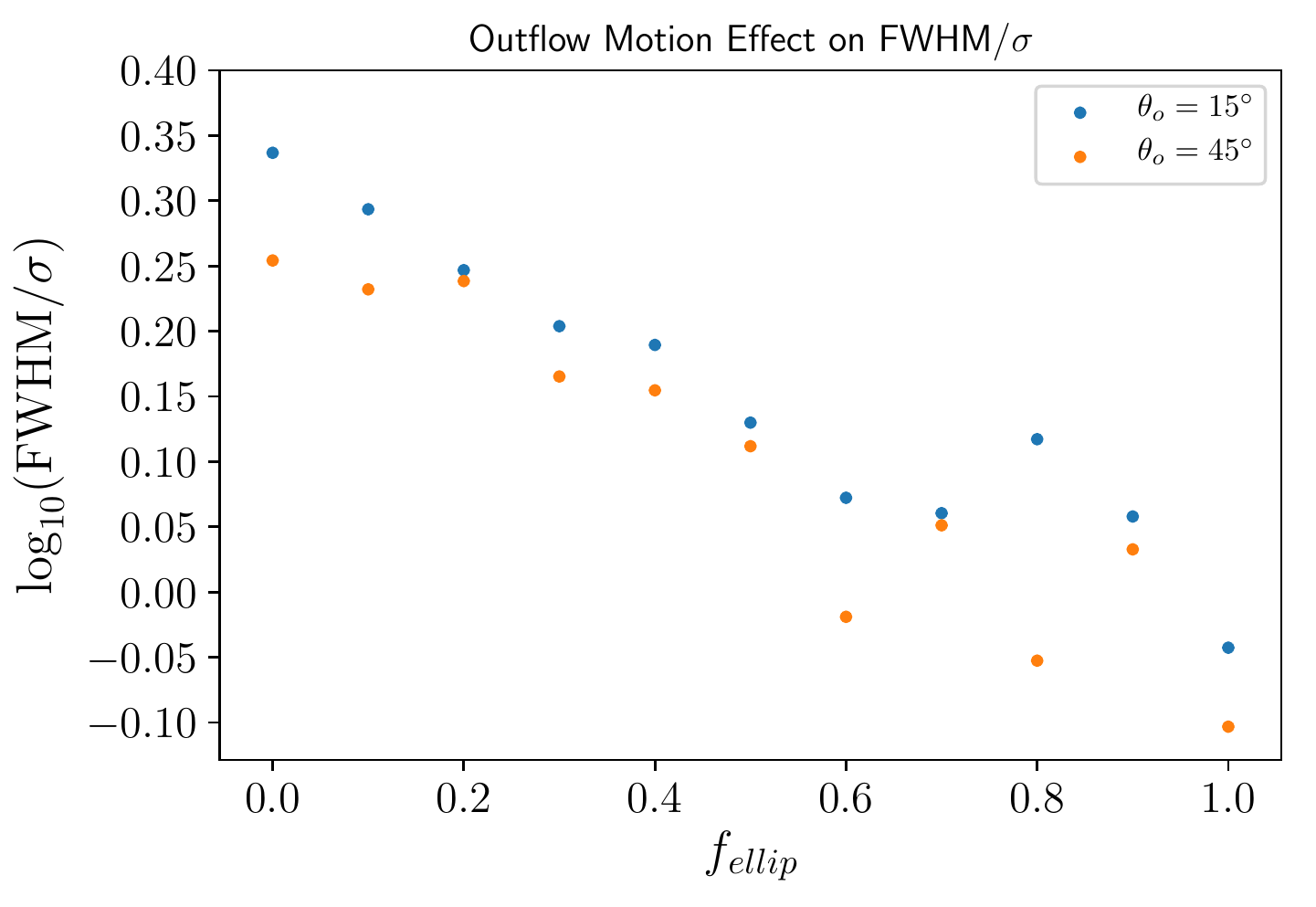}{0.5\textwidth}{}}
\caption{Inflow (left) and Outflow (right) effects on line profile shape.
Two different BLR disk thickness/opening angles, $\theta_o$ are used. In both plots, a thick disk with $\theta_o=15^{\circ}$ is shown in blue and a spherical structure with $\theta_o=45^{\circ}$ is shown in orange. The \textit{x}-axis, $f_{\rm{ellip}}$, represents the fraction of particles on elliptical orbits. Thus an increasing value of $f_{\rm{ellip}}$ represents a greater percentage of particles on elliptical orbits, rather than on radially inflowing/outflowing orbits. For both inflowing/outflowing motion, we see that line profiles with smaller values of $\log_{10}(\rm{FWHM}/\sigma)_{\rm{mean}}$ are produced with most of the particles on elliptical orbits with some inflow/outflow motion. Additionally, our results reconfirm our finding with thick diskness, a more spherical BLR produces broad lines with smaller values of $\log_{10}(\rm{FWHM}/\sigma)_{\rm{mean}}$, and confirm that inflowing/outflowing BLR motion is able to produce the line profile shapes we are particularly interested in, e.g. $\log_{10}(\rm{FWHM}/\sigma) < 0.2$. }
\label{fig: fig: blrkinematics_lineprofile}
\end{figure*}

\subsection{Inflow/Outflow Motion}
For this test we vary $f_{\rm{ellip}}$ (the fraction of particles with elliptical orbits) while keeping all other parameters held constant. The values selected for all other parameters are as follows:  $\theta_i=25^{\circ}$, 
$\beta=1.0$, $\mu=1$, $\log_{10}($\mbh$/M_{\odot})=7.5$, 
$\theta_{e}=45^{\circ}$, $\sigma_{\rm{turb}}=0.001$. We use a value of $f_{\rm{flow}}=0.1$ for inflow motion, and a value of $f_{\rm{flow}}=0.9$ for outflow motion. Additionally, we also use two separate disk thickness parameters for our test, $\theta_o=15^{\circ}$ and $\theta_o=45^{\circ}$. We remind the reader that $\theta_e=45^{\circ}$ and $f_{\rm{ellip}}$ represent particles on highly elliptical bound orbits (with  $1-f_{\rm{ellip}}$ on inflowing/outflowing orbits, as determined by the value of $f_{\rm{flow}}$). Therefore, a greater value of $f_{\rm{ellip}}$ represents a greater fraction of particles on elliptical orbits, rather than radially inflowing/outflowing orbits.

We find that inflowing/outflowing kinematics are able to produce broad-line profiles with smaller values of $\log_{10}(\rm{FWHM}/\sigma)_{\rm{mean}}$, i.e., $\log_{10}(\rm{FWHM}/\sigma) \approx 0.1$--0.2. In some cases, even values corresponding to $\log_{10}(\rm{FWHM}/\sigma) < 0.1$ are produced (see Figure \ref{fig: fig: blrkinematics_lineprofile}). These results also validate our previous finding in which flatter structures (e.g., $\theta_o=15^{\circ}$) produce broad lines with larger values of $\log_{10}(\rm{FWHM}/\sigma)_{\rm{mean}}$, compared to more spherical structures (e.g., $\theta_o=45^{\circ}$), which produce broad lines with smaller values of $\log_{10}(\rm{FWHM}/\sigma)_{\rm{mean}}$. We also see that $\log_{10}(\rm{FWHM}/\sigma)$ decreases, with increasing values of $f_{\rm{ellip}}$. A value of $f_{\rm{ellip}}=0.4$ corresponds to 40\% of particles on elliptical orbits, with the remaining 60\% on inflowing/outflowing orbits near escape velocity. While a value of $f_{\rm{ellip}}=0.9$ corresponds to 90\% of particles on elliptical orbits, with the remaining 10\% on inflowing/outflowing orbits. 
This suggests that a combination of inflow/outflow motion and highly elliptical orbits produces broad lines with smaller values of $\log_{10}(\rm{FWHM}/\sigma)_{\rm{mean}}$, rather than pure inflow/outflow motion.

\subsection{Turbulent Motion}
In addition to inflow/outflow motion, turbulence has also been suggested to cause the extended wings found in a Lorentz profile \citep{2013A&A...549A.100K}. We test the effect of turbulent motion on line-profile shape using the \textsc{caramel} model parameter $\sigma_{\rm{turb}}$, which allows for macroturbulent velocities. Since the random macroturbulent velocity that is added to the line-of-sight velocity of the particles, depends on both  $\sigma_{\rm{turb}}$ and $|v_{\rm{circ}}|$, we test with two different values of $\log_{10}($\mbh$/M_{\odot})$, as a larger black hole mass would result in greater magnitudes of circular velocity, and thus larger random macroturbulent velocities. Hence, we expect a more massive black hole, with greater turbulent motion, to have broad lines with smaller values of $\log_{10}(\rm{FWHM}/\sigma)_{\rm{mean}}$.

\begin{figure}[h]
\centering
\includegraphics[height=5.9cm,keepaspectratio]{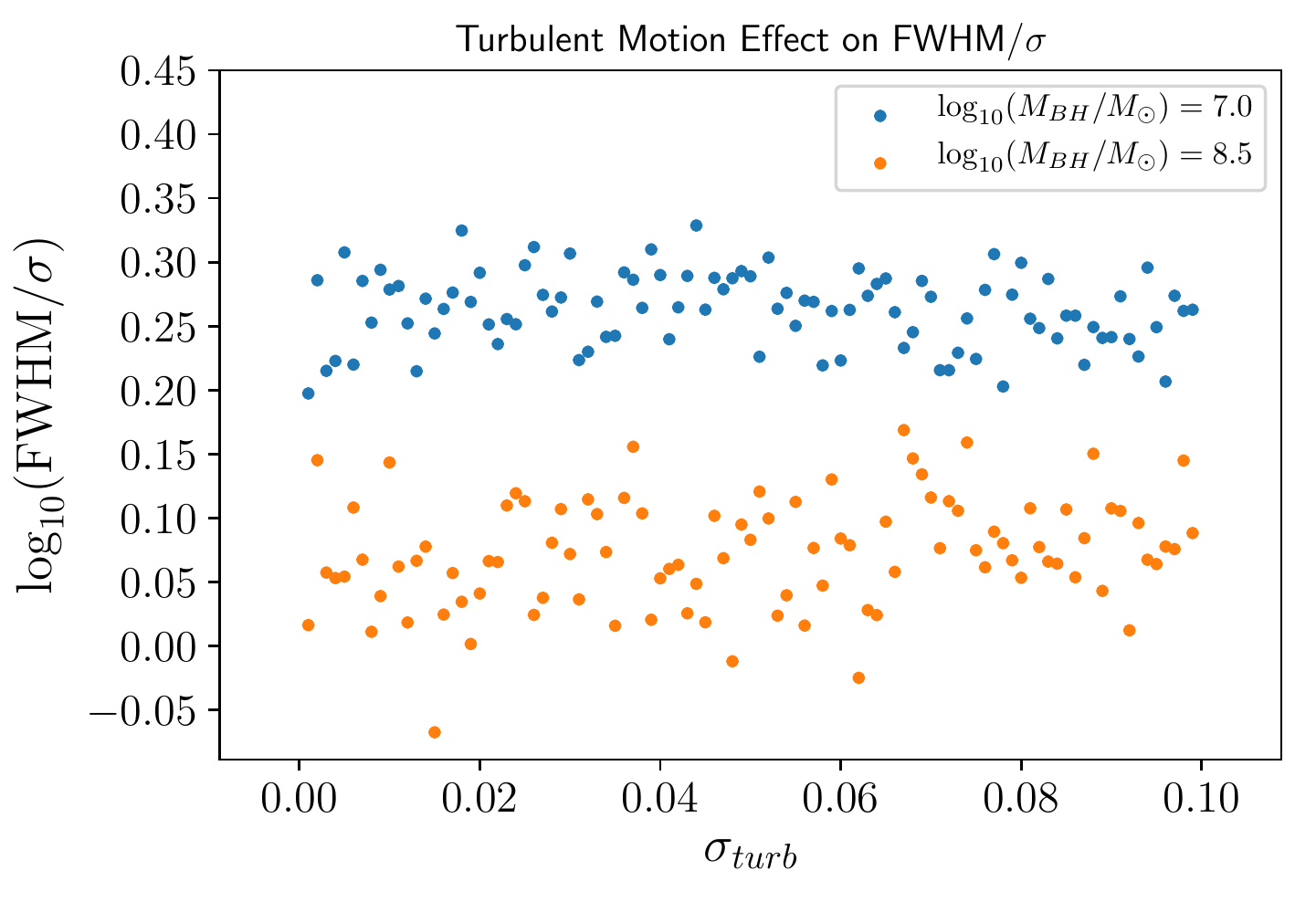}
\caption{We investigate the role of turbulent motion in line profile shape using \textsc{caramel} toy models by varying the parameter $\sigma_{\rm{turb}}$, and holding all other model parameters constant. Since macroturbulent velocities depend on both $\sigma_{\rm{turb}}$ and $|v_{\rm{circ}}|\propto \log_{10}($\mbh$/M_{\odot})$, we test the effects of turbulent motion using two different black hole masses. The blue points correspond to $\log_{10}($\mbh$/M_{\odot})=7.0$ and the orange points correspond to $\log_{10}($\mbh$/M_{\odot})=8.5$. As expected, we see the more massive black hole, which represents greater turbulent motion, produces broad lines with smaller values of $\log_{10}(\rm{FWHM}/\sigma)_{\rm{mean}}$.} 
\label{fig: blrturbulence_lineprofile}
\end{figure}

For both scenarios,  $\log_{10}($\mbh$/M_{\odot})=7.0$ and $\log_{10}($\mbh$/M_{\odot})=8.5$, we set particles on mostly bound outflowing orbits ($\theta_e=45^{\circ}$ and $f_{\rm{flow}}=0.9$) while varying the $\sigma_{turb}$ parameter within the limits of its prior, 0.001-0.1. As expected, we find the more massive black hole, $\log_{10}($\mbh$/M_{\odot})=8.5$, produces broad lines with smaller values of $\log_{10}(\rm{FWHM}/\sigma)_{\rm{mean}}$, with increasing macroturbulent contributions (see Figure \ref{fig: blrturbulence_lineprofile}).

\section{Conclusions}\label{sec: summary}
We use the direct modeling results of a sample of 28 AGNs --- nine from LAMP 2016 \citepalias{Villafa_a_2022}, seven from LAMP 2011 \citepalias{2018ApJ...866...75W}, four from AGN10 \citepalias{Grier++17}, five from LAMP 2008 \citepalias{pancoast14b}, one from AGNSTORM \citepalias{2020ApJ...902...74W}, one from \citetalias{bentz2021detailed}, and one from \citetalias{2022ApJ...934..168B}. The \textsc{caramel} results of these 28 AGNs provide insight into BLR geometry and kinematics, and constrain \mbh\ measurements without implementing the scale factor $f$ used in reverberation mapping estimates. The cross-correlation time lags and line widths reported by each subsample's respective reverberation mapping paper are employed to determine individual scale factors for each source. Using the extended sample described above, we search for existing correlations between scale factor and other AGN/BLR parameters/observables. Our main results are as follows.
\begin{enumerate}[(i)]
    \item We find $3.4\sigma$ evidence for a correlation between \logfmeansigma\ and black hole mass. 
    \item We find $2.8\sigma$ marginal evidence for a correlation between \logfrmssigma\ and black hole mass.
    \item We find $2.1\sigma$ marginal evidence for an anti-correlation between \logfmeanfwhm\ and BLR disk thickness.
    \item We find $2.4\sigma$ marginal evidence for an anti-correlation between \logfrmsfwhm\ and BLR disk thickness.
    \item We find $2.6\sigma$ marginal evidence for an anti-correlation between \logfmeanfwhm\ and BLR inclination angle.
    \item We find $2.4\sigma$ marginal evidence for an anti-correlation between \logfrmssigma\ and BLR inclination angle.
    \item We find $2.0\sigma$ marginal evidence for an anti-correlation between \logfmeansigma\ and BLR inclination angle.
    \item We find $2.2\sigma$ marginal evidence for a correlation between line profile shape measured from the rms spectrum, \lineprofile, and \logfrmssigma.
    \item We investigate how BLR properties may effect measured line profile shape using \textsc{caramel} models, and find that smaller BLR size, spherical geometries, inflow/outflow motion, and turbulent motion produce broad lines with smaller values of $\log_{10}(\rm{FWHM}/\sigma)_{\rm{mean}}$. 
    \item We conclude that these geometric \& kinematic effects cause a larger observed sigma line width (and cuspier FWHM/$\sigma$) at fixed \mbh, requiring a smaller virial factor, $f$, for black hole mass estimators.
\end{enumerate}
The sources modeled by \citetalias{Villafa_a_2022} have increased the number of AGNs with dynamical modelling of the BLR by nearly 50$\%$, and led to the discovery of a correlation with the scale factor and line-profile shape. The correlation with line-profile shape 
may provide an observational proxy for the virial coefficient in the future, however, further \textsc{caramel} studies and a larger sample are needed to confirm the statistical significance of the correlation. 

\begin{acknowledgements}

We thank the anonymous referee for their helpful comments and suggestions, which significantly improved this manuscript. L.V., P.R.W., and T.T. acknowledge support from the NSF through grant AST-1907208, ``Collaborative Research: Establishing the Foundations of Black Hole Mass Measurements of AGN across Cosmic Time.''

Research at UC Irvine has been supported by NSF grants AST-1907290. V.U acknowledges funding support from the University of California Riverside's Chancellor's Postdoctoral Fellowship and NASA Astrophysics Data Analysis Program Grant \#80NSSC20K0450. T.T. acknowledges support by the Packard Foundation through a Packard research fellowship. V.N.B. gratefully acknowledges assistance from NSF Research at Undergraduate Institutions (RUI) grants AST-1312296 and AST-1909297. Note that findings and conclusions do not necessarily represent views of the NSF. D.C.L.'s group at San Diego State University gratefully acknowledges support from the NSF through grants AST-1210311 and AST-2010001.

M.C.B. gratefully acknowledges support from the NSF through grant AST-2009230. G.C. acknowledges NSF support under grant AST-1817233. J.H.W. acknowledges funding from the Basic Science Research Program through the National Research Foundation of Korean Government (NRF-2021R1A2C3008486). A.V.F.'s group at U.C. Berkeley is grateful for support from the TABASGO Foundation, the Christopher R. Redlich Fund, the Miller Institute for Basic Research in Science (in which he was a Miller Senior Fellow), and many individual donors. 
We acknowledge the generous support of Marc J. Staley, whose fellowship partly funded B.E.S. whilst contributing to the work presented here in as a graduate student. 

This research has made use of the NASA/IPAC Extragalactic Database (NED), which is operated by the Jet Propulsion Laboratory, California Institute of Technology, under contract with the National Aeronautics and Space Administration. 
\end{acknowledgements}
\begin{appendix}
\section{\textsc{caramel} code modifications}
\label{caramel modifications}
The minor modifications made to the original \textsc{caramel} code, after its use by \citetalias{2018ApJ...866...75W}, \citetalias{Grier++17}, \citetalias{pancoast14b}, \citetalias{2020ApJ...902...74W}, \citetalias{bentz2021detailed}, and \citetalias{2022ApJ...934..168B}, (and prior to the use by \citetalias{Villafa_a_2022}) are outlined in the Appendix of \citet{2022ApJ...935..128W}. Here we summarize the content found in \citet{2022ApJ...935..128W}.

The original \textsc{caramel} model used by \citetalias{2018ApJ...866...75W}, \citetalias{Grier++17}, \citetalias{pancoast14b}, \citetalias{2020ApJ...902...74W}, \citetalias{bentz2021detailed}, and \citetalias{2022ApJ...934..168B} first draws the particles' radii from a shifted gamma distribution as described in the text. Then the particles are placed on the positive \textit{x}-axis and each particle is rotated around the \textit{z}-axis, by an angle drawn from a uniform distribution between 0 and 2$\pi$. The particles are then rotated about the \textit{y}-axis, by an angle drawn from the following distribution: $\arccos{(\cos{\theta_o}+(1-\cos{\theta_o})\times U^{\gamma}})$, where $\theta_o$ determines the opening angle (disk thickness) of the BLR and $U$ is a uniform distribution defined between 0 and 1. Additionally, in this original version of the code, $\gamma$ is allowed to range from 1 to 5. Upon this second rotation, the particles are rotated twice more --- once about the \textit{z}-axis by an angle drawn from a uniform distribution between 0 and $2\pi$ (which creates the thick disk), and once more about the \textit{y}-axis by an angle defined by $\pi-\theta_i$.

Prior to the \textsc{caramel} modeling of the LAMP 2016 sample, our team discovered that the second rotation about the \textit{z}-axis redacted the effect of $\gamma$ and modified the \textsc{caramel} code to allow for the effects of $\gamma$. The modified version of \textsc{caramel} used by \citetalias{Villafa_a_2022} varies from the placement of particles from the shifted gamma distribution. Rather than place all particles on the positive \textit{x}-axis as described above, particles are placed on both positive and negative sides of the \textit{x}-axis. Then the particles are only rotated a total of three times, rather than four. The first rotation is about the \textit{y}-axis, rather than the \textit{z}-axis, and is defined by an angle drawn from the following distribution: $\arcsin{(\sin{\theta_o}\times U^{1/\gamma}})$, which creates a double wedge in the \textit{xz} plane. After the first rotation, the particles are then rotated about the \textit{z}-axis by an angle drawn from a uniform distribution between 0 and $2\pi$ (which creates a thick disk). Then, the particles are rotated by one final rotation about the \textit{y}-axis by an angle defined by $\pi-\theta_i$.

After the changes made in geometric construction, we noticed that most of the effects of $\gamma$ occur within the ranges $\gamma=1$--2, and changed the priors on the parameter accordingly.
\end{appendix}
\bibliographystyle{apj}
\bibliography{references}
\end{document}